# High-efficiency atmospheric water harvesting enabled by ultrasonic extraction

Ikra Shuvo[1], Carlos D. Díaz-Marín[2], Marvin Christen[4], Michael Lherbette[4], Christopher Liem[3], Svetlana V. Boriskina[2,*]

[1]*Media Lab, Massachusetts Institute of Technology, Cambridge, MA 02139, USA*
[2]*Department of Mechanical Engineering, Massachusetts Institute of Technology, Cambridge, MA 02139, USA*
[3]*Department of Electrical Engineering and Computer Science, Massachusetts Institute of Technology, Cambridge, MA 02139, USA*
[4]*SmarAct Metrology GmbH & Co. KG, Oldenburg, HRA 207391, Germany.*
*Corresponding author: <u>sborisk@mit.edu</u>

## Abstract

Atmospheric water harvesting technology, which extracts moisture from ambient air to generate water, is a promising strategy to realize decentralized water production. However, the prohibitively high energy consumption of heat-induced evaporation process of water extraction hinders the technology deployment. Here we demonstrate that vibrational mechanical actuation can be used instead of heat to extract water from moisture harvesting materials, offering about forty-five-fold increase in the extraction energy efficiency.  We report the energy consumption for water extraction below the enthalpy of water evaporation, thus breaking the thermal limit of the energy efficiency inherent to the state-of-the-art thermal evaporation and making atmospheric water harvesting technology economically feasible for adoption on scale.

## Introduction

Many communities worldwide are located in arid climate zones and experience shortage of fresh water resources, which severely inhibits their land development and creates harsh humanitarian conditions for their populations [1]. More than half of Mexico's national territory lies in the arid and semi-arid climate zones, including the Chihuahuan and Sonoran deserts, which extend northward into the United States. The Colorado river is drying up at an alarming rate, putting these areas at even higher risk of freshwater shortages. Similarly, in southern Ukraine, desertification of thousands of hectares of fertile steppe land may occur following the breach of the Kakhovka Dam by the Russian army in 2023, with potentially grave consequences for the world food supply. The situation is even more dire in the Middle East, where only three major rivers supply the vast region with fresh water, and where most land areas are arid. In many cases, the climate problems are exacerbated by the economic and security-related ones, which make the costs of the industrial installations for water purification prohibitive for communities in need [2,3].

Fortunately, the Earth atmosphere provides a vast resource of fresh water. Fog harvesting has been ubiquitously practiced by plants, animals, and humans throughout history [4–10]. However, the majority of the 12,900 billion tons of fresh water stored in the atmosphere [11] is not available as fog, which only forms when the relative humidity (RH) of air approaches 100%. Accordingly, there is a need for atmospheric water harvesting (AWH) technologies that can operate in arid or semi-arid climates or in regions where large-scale installations are impractical by economic or security reasons[12–18]. These typically fall into two major categories, including (i) active refrigeration technology[19,20] to cool air below the dew point and promote condensation[20–23], and (ii) sorption-desorption technology that employ porous hydrophilic sorbents to collect moisture from air, followed by water extraction via the heating-induced evaporation-condensation process[12,21,24,25].

However, the former technology is hard to scale down to realize affordable decentralized water production, while the latter one in its current state of development exhibits prohibitively high energy consumption associated with the heat-driven process of water desorption from AWH materials. The temperature needed to release the captured water from a sorbent can be as high as 160°C for conventional desiccants such as silica gel, zeolite, and activated alumina[26–28]. Desorption temperatures of other recently-developed sorbents including hydrogels[15], metal-organic frameworks (MOFs)[12,21], anhydrous and hydrated salt couples[13] are in the 60 to 80°C range, which makes water release an energy-intensive process[29,30]. Furthermore, all the AWH prototypes demonstrated to date exhibit at least an order of magnitude higher energy consumption than the predicted thermodynamic limit of about 2MJ/kg at 30% RH[29]. This is a major bottleneck of the sorption-desorption AWH technology, which limits its contribution to the United Nation's 6th Sustainable Development Goals (SDG6)[31].

The ongoing efforts to increase water harvesting capability and reduce the energy consumption mostly focus either on engineering more efficient sorbents (i.e., capable of sorbing higher weight of water per gram of material)[32,33] or/and on designing autonomous AWH systems that can utilize solar energy[12,13,21,33]. Since the energy consumption occurs during the desorption process, finding alternative technologies to replace heat-driven evaporation is the key to achieving a breakthrough in the AWH technology feasibility. Recent studies showed that re-engineering the sorbent structure to facilitate water evaporation at lower temperatures, e.g., by incorporating a secondary network of reconfigurable phase-change polymers[30], or by combining heating with other external stimuli such as sunlight or laser illumination can increase the evaporation efficiency, even beyond the thermodynamic limit[34].

Here, we propose a new method of moisture extraction, which uses vibrational mechanical actuation enabled by piezoelectric materials (**Fig. 1,** Supplementary Movie 1, Supplementary Note 1, Supplementary Fig. S1) to extract water from AWH sorbents. We draw inspiration from prior studies that used of piezoelectric actuators[35] and Si-based Micro-Electro-Mechanical-system (MEMS)[36] for atomization of liquids. We show that this method offers at least a five-fold reduction of the energy needed to extract the same mass of water as the standard evaporation-condensation technique. While demonstrated here for hydrogels, the technique is very general and may be used with other sorbents, including MOFs, superabsorbent fibers and textiles, salts, and desiccants. Since the new technology is not based on the heating-induced evaporation, its efficiency is not bound by the thermodynamic limit derived for the conventional desorption process, offering promise for further efficiency improvements.

## Results

**Figure 1a** illustrates a piezoelectric-ceramic-based single-head ultrasonic actuator used in this study for water extraction from several atmospheric water harvesting hydrogels (AWH-Hs). High-voltage-driven lead zirconate titanate (PZT) ring converts applied electrical power into the in-plane displacement of the ring, which in turn causes out-of-plane mechanical oscillation of the steel mesh membrane clamped to the ring. Mechanical oscillation of the membrane is accompanied by the Joule heating caused by ferroelectric hysteresis loss of the PZT material when driven at high frequency,[37] as illustrated in **Fig. 1b**. Here, we show (for the first time to the best of our knowledge) that a synergetic impact of the mechanical actuation and device heating facilitates moisture extraction from AWH-Hs, reduces the extraction time, and increases the system energy efficiency. The first prototype device utilizing this water extraction method is shown in **Figs. 1c-d**.

In **Fig. 1e**, we plot the energy efficiency of water extraction $\eta$, which is defined as a ratio of the enthalpy of water vaporization $h_{lv}$ ($h_{lv} = 2.257\ MJ/\text{kg}$ at 100°C) and the energy consumption of the device needed for the extraction of a unit mass of water from a sorbent, $E$ [MJ/kg]: $\eta = h_{lv}/E$ [29,38,39] (see Supplementary Note 4 and Supplementary Table 4). In an ideal thermal process, where the enthalpy of desorption is identical to the enthalpy of vaporization of water, all the heat provided to the system is used for water evaporation, i.e., $E = h_{lv}$. In this work, we refer to this ideal thermal process with $\eta = 100\%$ as the thermal limit. In any real thermal water release process, the heat provided is used not only evaporate water, but also to sensibly heat the system and sorbent, and is partially lost to the ambient, leading to $\eta < 100\%$. Conversely, an efficient *non-thermal* extraction method (such as, e.g., vibrational actuation) can have a specific energy consumption lower than the enthalpy of vaporization of water, $h_{lv}$. Then, the energy efficiency of the water release process can in principle exceed 100%.

**Figure 1e** compares the efficiency of our device operated with a polyacrylamide and lithium chloride (PAM-LiCl) hydrogel sorbent with that of the-state-of-the-art AWH devices[40–44]. The extraction efficiency values plotted in **Fig. 1e** for different systems account for the specific heat of both water and other solids used in the system (i.e., sorbent materials, heaters, enclosures, etc.) as well as heat dissipation losses. Each of the devices in **Fig. 1e** represents a single-stage system without heat recovery, and thus these efficiency values are all less than 100%. Our device shown in **Fig. 1a-d** demonstrated energy efficiency of almost 428%, exceeding the state-of-the-art efficiency of 9.5%[39,45] by a factor of ~45 (**Fig. 1e**). In the following, we analyze the performance of the system operated with hydrogel-based sorbents, but the concept is much broader, encompassing a variety of AWH materials and actuator types, and showing a potential to further increase the efficiency.

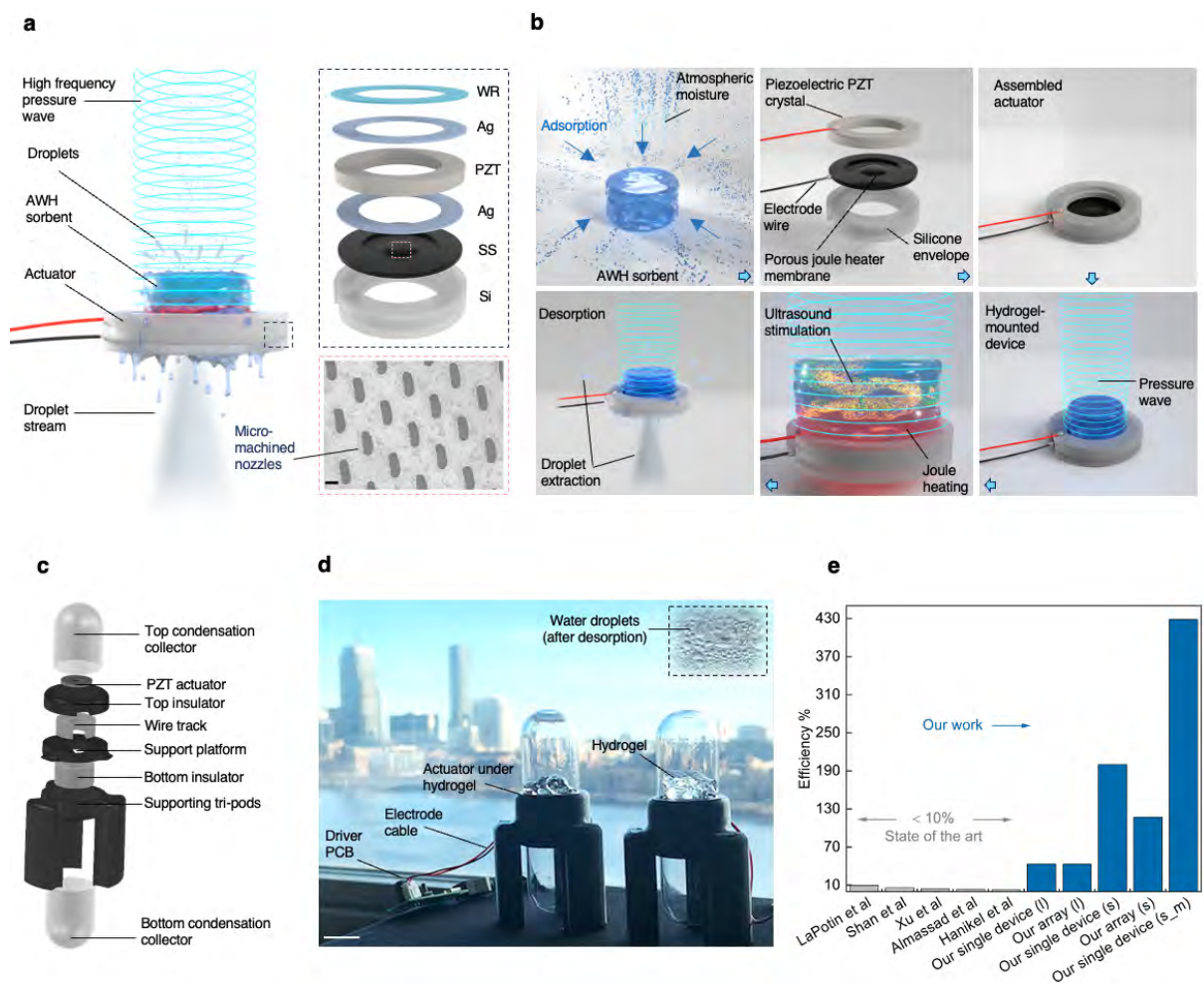

Figure 1: Ultrasonic moisture extraction concept and a high-efficiency extractor prototype. **a.** The ultrasonic extractor comprises a PZT-crystal piezoelectric transducer and a stainless-steel (SS) porous membrane through which the desorbed water is extracted from a sorbent material under vibrational actuation. The black dashed box shows the transducer structure, including an Ag-coated PZT ring topped with a thin layer of water-resistant resin attached to the SS membrane and encased in a silicone elastomeric ring. The red dashed box shows the structure of the micro-machined nozzles on the SS membrane, which assist in directing the flow of water out of the device. **b.** The diagram of the system assembly and sorption/desorption stages of the moisture harvesting process utilizing a sorbent material and a micro-actuator. The top Ag layer and the SS membrane are connected to the electrode wires. **c.** A CAD model of the system custom-designed to collect the moisture extracted from the sorbent by the actuator. **d.** A photograph of the FDM-printed device prototypes with hydrogel sorbents in the process of moisture harvesting from ambient air. The inset shows a close-up view of the collected water droplets on the glass enclosure. **e.** Efficiency values of the moisture extraction achieved in this study compared with the state-of-the-art literature data (here, s = short actuation period; l= long actuation period; s_m = multiple repetitive cycles of short actuation periods). Scale bars: 50 μm (a); 15 mm (d).

We synthesized three types of PAM-LiCl atmospheric water harvesting hydrogels (see Methods), which incorporate lithium ($Li^+$) and chloride ions ($Cl^-$) as essential elements for their ability to adsorb water molecules from atmosphere (**Fig. 2a**)[46]. Hydrogels had the same salt and polymer content, but different amounts of N,N'-methylenebisacrylamide (MBA) crosslinker, and thus exhibited different crosslinking densities. Hydrogels labeled HG-A, HG-B, and HG-C had crosslinker-to-monomer mass ratios of 0.1, 0.5, and 10 mg/g, respectively. HG-A had very low rigidity, while HG-C could stand on its own outside a Petri dish without collapsing (**Fig. 2b**). As a result, HG-A hydrogel demonstrated greater resistance to tearing when manually stretched, in contrast to the more rigid HG-C, which can be torn easily (**Fig. 2c** and Supplementary Movie 2). The differences in the hydrogels micro-structure and surface morphology are revealed by the scanning electron microscope (SEM) images, with HG-A exhibiting a low-density structure (**Fig. 2d**) and HG-C (**Fig. 2e**) – a more compact one. We also included in our study a commercial hydrogel (a cross-linked glycerol- 2-Acrylamido-2-methylpropanesulfonic Acid sodium salt hydrogel), and labeled it HG-M.

The four types of AWH-Hs were compared by their elastic storage moduli (see Methods), which provide a measure how much energy is stored in the material when it is subjected to a periodically oscillating load, and their loss moduli, which measure the material's ability to dissipate applied stress through heat (**Fig. 2f** and Supplementary Figs. S2-S3). The progressively higher storage moduli of hydrogels with higher MBA content validate the strategy of engineering sorbents with desired mechanical properties by varying the density of chemical crosslinks[47]. All the PAM-LiCl sorbents exhibited loss moduli lower than their storage moduli, and thus the resultant phase shift between the mechanical deformation and the sorbent material response, $\tan \delta = G''/G' < 1$. This indicates the elasticity and ability of these materials to bounce back to their original shape as long as the applied deformation is below their yield point[48]. These data are similar to those previously reported for salt-free PAM hydrogels[49], suggesting that the high salt content does not play a major role in their mechanical properties, which are mostly controlled by the polymer crosslinking density. In turn, the commercial hydrogel (HG-M) exhibited higher values of storage and loss moduli as well as $\tan \delta$ value, thus demonstrating higher mechanical strength than PAM-LiCl AWH-Hs, but slightly lower elasticity.

**Figure 2g** shows that all three PAM-LiCl AWH-Hs exhibit high water uptake in different environments, from arid to very humid, between 15% and 85% relative humidity, RH (see Supplementary Movie 3 for water harvesting at ~15% RH). This is consistent with recent reports on PAM-based hydrogels and hygroscopic PAM-LiCl composite hydrogels exhibiting a large capability to capture moisture[30,32]. Water harvesting capacity at the same RH level is similar for all the PAM-LiCl AWH-Hs regardless of their crosslinking density[32,50,51]. In contrast, the commercial HG-M did not demonstrate any water sorption capability at low RH (15% and 30%).

The goal of the study is to use a piezoelectric actuator to generate and transmit a pressure wave into the AWG hydrogels to facilitate water extraction. To optimize the actuator geometry and performance, several candidate devices were evaluated. Based on prior literature for water atomization, we targeted a resonance frequency in the 100-160 kHz range[33,35,36,52,53]. Four piezoelectric ultrasound transducers with resonance frequencies ranging from ~107-115 kHz to ~165 KHz were designed and analyzed (Supplementary Fig. S4). Three devices had similar geometry but differed by the size of micro-nozzles perforated in their metallic membranes. The fourth device (PZ-D) had a comparatively lower diameter to thickness aspect ratio, ensuring higher (~165KHz) resonance frequency (see Supplementary Table 1). All the four devices demonstrated comparable piezoelectric coefficients ($d_{33} = ~435 - 440$ pC/N), which were measured using "quasi-static" or "Berlincourt" method, employing a static force sensor and piezoelectric meter (see Methods and Supplementary Fig. S5). The piezoelectric charge coefficient, denoted as $d_{33}$, is determined by applying a unit stress in the direction 3 (i.e., z-axis) that induces polarization in direction 3 (i.e., in the same direction as the polarization of the ceramic element).

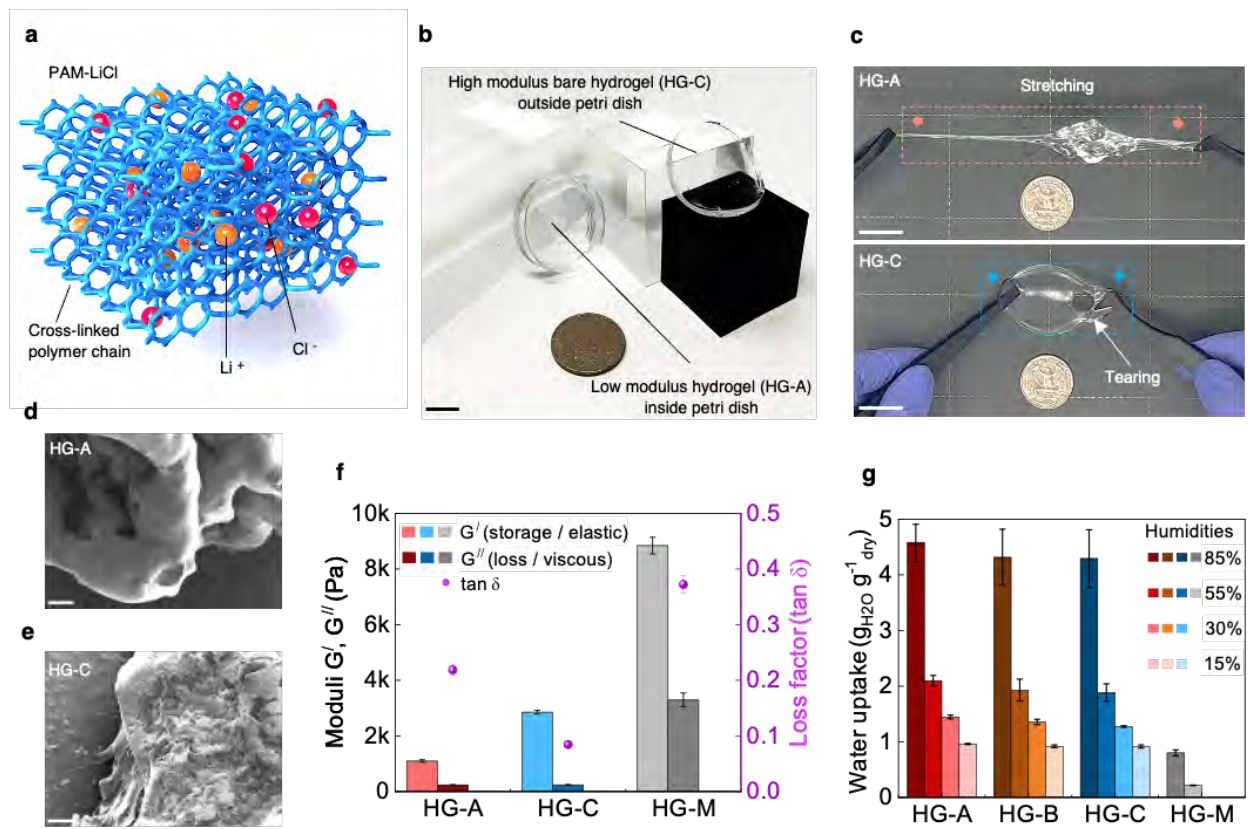

**Figure 2. Morphology, water harvesting and rheological properties of AWH-Hs. a.** A schematic of the structure of a PAM-LiCl hydrogel with solvated lithium (Li⁺) and chloride (Cl⁻) ions diffused into a porous polymer network. **b.** Photographs of synthesized PAM-LiCl hydrogels inside and outside a Petri dish, illustrating the contrast between the samples with the highest (HG-A) and the lowest (HG-C) storage moduli. **c.** Optical images of the HG-A and HG-C hydrogels under manual stretching. **d,e.** SEM images of HG-A (d) and HG-C (e) samples. **f.** Mechanical properties of hydrogels measured at 25°C and with an oscillatory load at 6.3 rad/s. **g.** Water uptake (g/g) of hydrogels at four different RH levels at 25°C. Scale bars: 12 mm (b); 24 mm (c); 200 μm (d); 300 μm (e).

All four devices were designed such that the diameter of the piezoelectric ceramic ring was at least ten times larger than its thickness ($d > 10t$) to maximize their displacement amplitude[54]. The efficiency of the piezoelectric devices in converting electric energy into mechanical energy was evaluated by their effective electromechanical coupling coefficients ($k_{eff}$)[55,56], which were found to be around 0.19 for the low-frequency devices and 0.17 for the high-frequency device (Supplementary Fig. S6). The rest of the electrical power is dissipated as heat, leading to some temperature increase of the actuator and AWH-H sorbents.

First, the nozzle size was varied to improve water extraction through the membrane without degrading the resonant behavior of the actuator, as resonance frequencies of piezoelectric devices are sensitive to variations in the membrane mass and stiffness[57]. PZ-A, PZ-B, and PZ-C devices were designed to each have 1000 nozzles with diameters of 10, 70, and 100 μm, respectively (Supplementary Fig. S7). **Fig. 3a** compares the water conductance through the membranes with different nozzle sizes measured using a forward-osmosis (FO) configuration (see Methods and Supplementary Fig. S8). The water transport rate through the PZ-C membrane was found to be the highest, owing to its larger nozzle size.

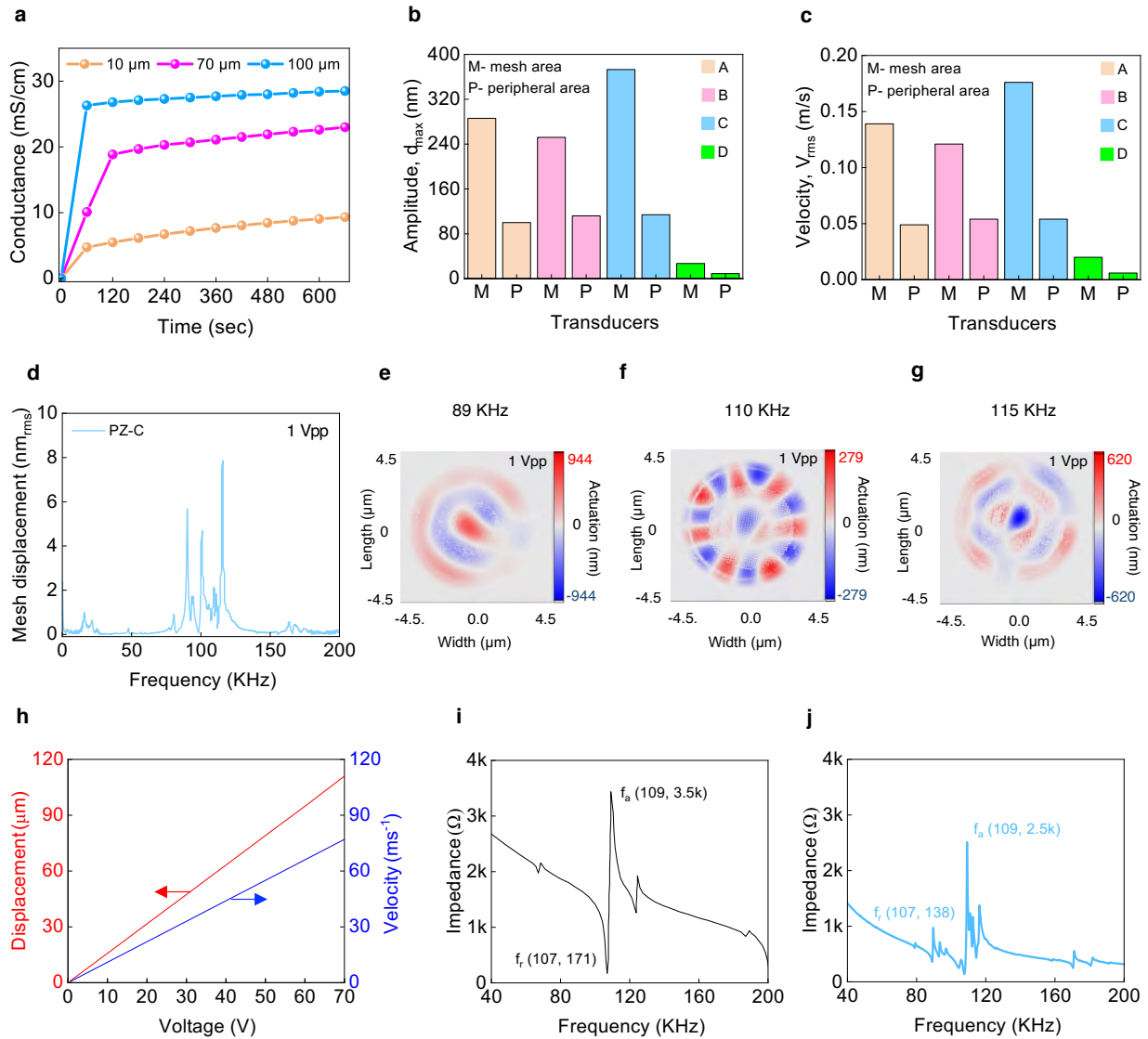

**Figure 3. Ultrasonic actuator selection and characterization. a.** The effect of the increased nozzle size on water conductivity through the porous membrane as determined by a forward-osmosis experiment. Here and in **b-c**, the data for membranes in PZ-A, PZ-B, and PZ-C transducers with nozzle sizes of 10 μm, 70 μm, and 100 μm are color-coded as orange, pink, and blue, respectively. PZ-D actuator had the same nozzles as PZ-A. **b-c.** Deflection amplitudes (b) and corresponding RMS velocities (c) of the four transducers subjected to a sinusoidal input signal at 1 Vpp measured at resonant frequencies of ~110 KHz for PZs A, B, and C, and ~165 kHz for PZ-D. **d.** Experimental modal analysis of the porous membrane attached to the ring of PZ-C transducer at 1 Vpp actuation. **e-g.** The deflection maps across the PZ-C membrane at frequencies of 110 KHz (e), 89 (f), and 115 KHz (g) measured by a PICOSCALE vibrometer under 1 Vpp (0.35 Vrms) sinusoidal signal input. **h.** Simulated displacement and velocities of the PZ-C membrane plate under higher voltages. **i-j.** The simulated (i) and measured (j) impedance spectra the PZ-C actuator.

AWH-Hs are actuated by the motion of the free-standing portion of the membrane that is not clamped to the PZT ring (**Fig. 1a**), and both the magnitude and velocity of its movement need to be maximized for efficient extraction. The diameters of these areas measured 7.8 mm for PZ-A, B, C actuators, and 5 mm for the PZ-D device. **Figure 3b** compares the maximum displacement amplitudes measured within the nozzle areas (M) of the membranes and the peripheral nozzle-free (P) areas, while **Fig. 3c** shows the corresponding RMS values of the deflection velocity. The transducers operating at lower frequencies (PZs-A, B, C) exhibit larger amplitude and velocity, both within the nozzle area and the periphery, compared to the higher-frequency device (PZ-D). For all the devices, the displacement amplitudes and velocities are higher in the center than on the periphery, as expected. PZ-C actuator membrane exhibited the largest displacements and RMS velocities. We attribute this to the larger nozzle diameter, which reduced the weight of the membrane, lowered its inertia, and modified its vibrational modes spectrum, leading to better coupling between the piezo oscillations and the membrane vibrations.

The resonant spectra of the membranes integrated in the actuators were measured by using a vibrometer at 1 Vpp (**Fig. 3d** and Supplementary Figs. S9-S10). Experimentally mapped vibrational mode profiles of the PZ-C membrane at its three dominant frequencies (89, 110, and 115 KHz) are shown in **Figs. 3e-g**. Supplementary Movies 4-5 and Supplementary Figs. S11-S12 visualize spatial amplitude and phase distributions of these vibrational modes. The actuation voltage of 1 Vpp was used for the above measurements to prevent the displacement amplitude over-range, as the vibrometer is not capable of detecting displacements larger than half-wavelength in magnitude ($\lambda/2 = 775\ nm$). However, during the desorption process, the device is actuated by a much larger voltage of ~114 Vpp using an amplifier with a high gain to boost the signal strength. To evaluate the displacement and velocity of the PZ-C actuator membrane at its operational voltage, we simulated the device performance via finite-element modeling with the COMSOL Multiphysics software (**Fig. 3h**). Modeling predicts that both the membrane displacement and the speed grow linearly with applied voltage, reaching the values of 111 μm and 77 m/s, respectively, at 70V.

The simulated spectrum of the PZ-C device (**Fig. 3i**) exhibits a characteristic Fano feature with the resonance (107 KHz) and anti-resonance (109 KHz) frequencies, which is in good agreement with the corresponding spectrum measured experimentally in the fabricated PZ-C device (**Fig. 3j**). The measured spectra for devices PZ-A, PZ-B and PZ-D are shown in Supplementary Fig. S13. The resonance frequency of device PZ-D was measured to be around 170±5 KHz.

Finally, our finite-element simulations of the PZ-C membrane oscillations (see Methods) reveal the vibrational mode pattern forming under the piezo actuation and predict the values of M displacement (790 nm) (Supplementary Fig. S14) and corresponding velocity (0.54 m/s) (Supplementary Fig. S15) at 1 Vpp, which is in reasonable agreement (within the order of magnitude) with their experimentally measured counterparts (i.e., 373 nm and 0.18 m/s, respectively).

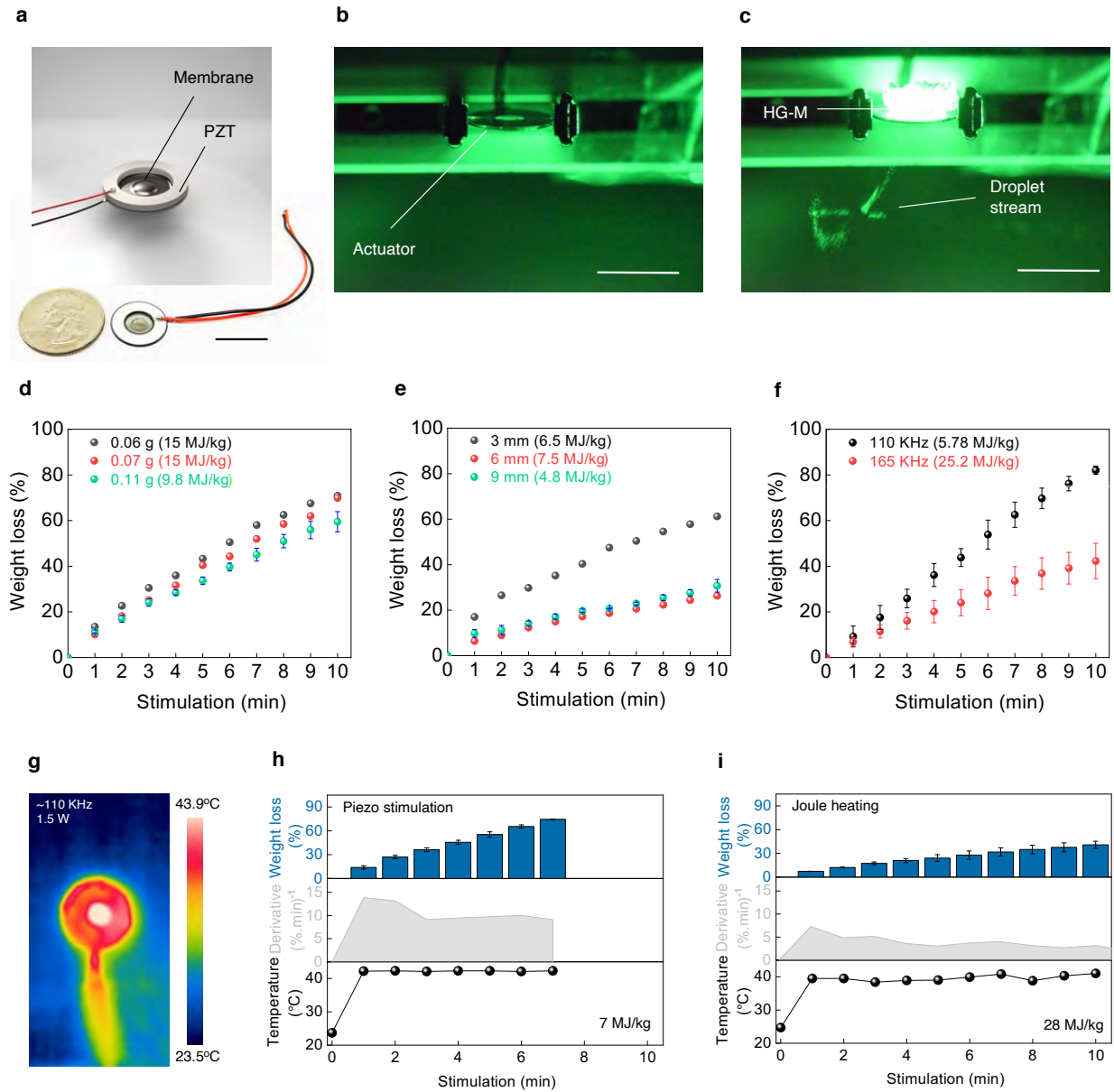

**Figure 4. Protocol development for water extraction with actuators driven at a constant 1.5 W power**. **a**. A schematic and a photograph of the actuator, comprising a steel membrane acting as a Joule heater and an ultrasound-generating piezoelectric component (PZT ring). **b-c**. Photographs of an HG-M sample on the actuator and of the water droplets ejecting under stimulation (scale bars 16 mm). **d**. Desorption kinetics measured as a function of the HG-M moisture content (quantified as a total initial weight of the specimen soaked in water). **e**. Desorption kinetics as a function of the sample geometry. **f**. Desorption kinetics as a function of the transducer actuation frequency. **g**. An IR image visualizing the Joule heating of the actuator operating at a 1.5 W power (applying ~114 Vpp) and a 110 KHz frequency. **h-i**. The rate of water extraction and membrane temperature under piezo actuation (**h**) and Joule heating (**i**). The energy consumption is calculated and listed for each curve in panels **d-f** and **h-i**.

**Figure 4a** illustrates an actuator, which includes a PZT ring and a membrane and acts both as a mechanical transducer and as a Joule heater. To analyze and optimize the extraction process, the AWH-G samples were soaked in the DI water and placed on top of the membrane (**Figs. 4b-c** and Supplementary Fig. S16). We first used a commercial hydrogel (HG-M) to evaluate the efficiency of water extraction and to establish an optimum experimental protocol of sorbent actuation. To visually capture the water extraction process, a green laser was focused across the path of the ejected water droplets, and photographs were captured using the burst mode of a digital camera with the shutter speed of $1/30^{th}$ of a second (**Figs. 4b-c**). These experiments revealed that – differently from the conventional evaporation-driven extraction process – water is released in the liquid form, although droplets large enough to be captured by our camera are only visually observable for only for a few seconds after the onset of piezo actuation.

We conducted four series of experiments to evaluate the role of the sorbent size, weight, moisture content, and nozzle size as well as the actuation frequency on the water extraction efficiency (**Figs. 4d-f** and Supplementary Fig. S17). First, HG-M samples of 3 mm diameter and 1 mm thickness were submerged in DI water for 30 s, 60 s, and 120 s, respectively. The samples were weighed before being positioned in the center of the mesh membrane of a PZ-A actuator (7.8 mm diameter and 10 μm nozzle size). The wet samples having initial weights of 0.06, 0.07 and 0.11 g, respectively, were exposed to piezoelectric stimulation for 10 minutes, and their weight loss is reported in **Fig. 4d** as a function of time. The heaviest sample exhibited the best performance in terms of energy consumption (9.8 MJ/kg), about 35% lower than 15 MJ/kg value measured for the samples with lower moisture content. We attribute the observed improvement to the better contact between the vibrating mesh membrane and a heavier specimen.

Second, we studied the effects of both sorbent mass and geometry by testing specimens of varying sizes under the same actuation protocol. Samples with diameters of 3, 6, and 9 mm and thicknesses of 1 mm were submerged in DI water for 120 s each, and their weight loss kinetics under actuation is reported in **Fig. 4e**. The largest and heaviest (~0.4 g loaded weight) sample exhibited the lowest energy consumption of 4.8 MJ/kg. The plots in **Figs. 4d-e** indicate that (i) the energy consumption per unit mass of extracted water does not always correlate positively with the percentage of the sample weight loss under actuation, (ii) higher efficiency is achieved when the sample fully covers the actuator membrane, and (iii) heavier samples with higher moisture content exhibit higher energy efficiency. This suggests that the optimum sample weight (achieved for a certain geometry and moisture content) needs to be found to minimize the energy consumption, as extra-heavy samples can dampen the membrane vibrations, reducing extraction efficiency.

Third, we compared the extraction efficiency achieved by using different actuator membrane geometries (Supplementary Fig. S17). As expected, the PZ-C actuator with the nozzle size of ~100 μm demonstrated lower energy consumption (7.7 MJ/kg) than PZ-A (16.6 MJ/kg), while PZ-B actuator with the nozzle size of ~70 μm had performance similar to that of PZ-C. The increased energy efficiency of PZ-C actuator can likely be attributed not only to the higher diffusion rates (**Fig. 3a**) but also to the larger membrane displacement and higher velocity (**Fig. 3b-c**).

Finally, piezo actuators operating at different frequencies, PZ-C (107 KHz) and PZ-D (165 KHz) with similar nozzle numbers and sizes, were used to extract moisture from HG-M samples of 3 mm diameter and 1 mm thickness, after they were submerged in DI water for 60 s. PZ-D actuator exhibited the highest energy consumption of 25.2 MJ/kg, indicating that the operating frequency plays a very important role in the water release process.

For all the experiments in **Figs. 4d-i**, the piezo actuator was driven by using a miniaturized and portable printed circuit board (PCB) electronic driver system at a constant ~70% duty cycle (defined as the ratio of the time during which the system is in an activated state (ON) and the overall duration of each cycle that comprises a series of ON-OFF cycles). Our modeling predicts that the axisymmetric displacement and velocity in the out-of-plane vibrations of the membrane at an applied voltage of 40 Vrms or ~114 Vpp reach ~60 μm and ~45 m/s, respectively (Supplementary Fig. S18). At the same time, infrared imaging (see Methods) reveals that the actuator can exhibit a temperature rise of >20°C within 60 s while working at 70% duty cycle and 1.5 W supplied power applying ~114 Vpp (**Fig. 4g**). It returns to the ambient temperature within a comparable time frame after actuation stops (Supplementary Fig. S19). This fast increase in temperature facilitates desorption of water from hydrogels[50], and the combined effect of fast temperature rise and hydrogel vibration synergistically drive quick and efficient water release.

The rate of water extraction and the energy consumption of the piezo-actuation and Joule heating methods are compared in **Figs. 4h-i**. Both the Joule heater (a bare stainless-steel membrane) and the piezo actuator were powered with 1.5 W, and the weight loss process of DI water-loaded HG-Ms was measured for a period of 10 min. The samples were actuated for 1-minute-long intervals and weighed after each minute, with both the weight loss and its derivative plotted in **Figs. 4h-i**. The bottom panels of **Figs. 4h-i** show the temperatures of the membranes measured by K-type thermocouple sensors, which were deliberately kept at the same level by choosing the 1-minute-long actuation periods (see Methods). The piezo-actuated specimen regained its original weight within a span of 7 minutes under piezo actuation at 42°C (**Fig. 4h**). In contrast, the sample subjected to waver evaporation via Joule heating did not regain its initial dry weight even after 10 minutes at 41°C (**Fig. 4i**). These data also revealed that the piezo-driven extraction consumed a significantly lower amount of energy (7 MJ/kg), a reduction of approximately 74% compared to the Joule-heater-induced evaporation (28 MJ/kg). A piezo actuation duty cycle can be further optimized to mitigate the generation of heat and increase efficiency[58].

It is important to ensure that the sorbet maintains its structural integrity under a combination of mechanical and heat stimuli during the water extraction process. The inspection of HG-M specimens and the piezo membrane after multiple tests showed no visible damage on the hydrogel and no significant material residue left on the membrane, as evidenced by the scanning electron microscopy (SEM) image (Supplementary Fig. S20). The very few nozzles that appeared jammed were easily cleared by rinsing the membrane with DI water, thereby enabling its repetitive utilization. Surface topography scans of the membrane also revealed a relatively even terrain (Supplementary Figs. S21-S22).

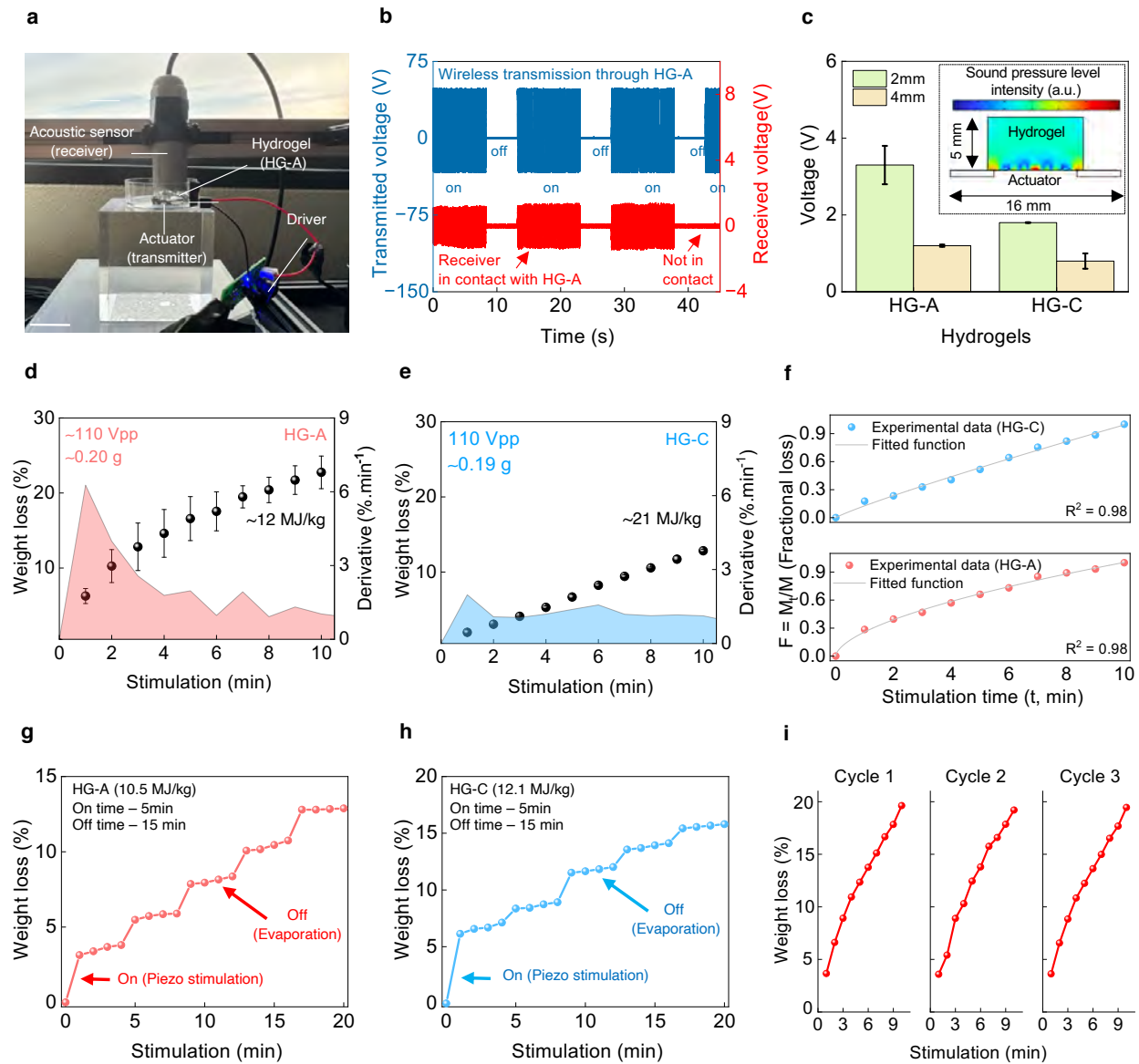

**Figure 5. The effect of AWH-H mass and stiffness on acoustic signal attenuation and energy efficiency of water extraction. a.** A setup for measuring transmittance and attenuation of acoustic waves propagating through hydrogel samples. A sample is placed between the actuator (transmitter) and a sensor (receiver), and the transmitted signal is measured as the output voltage. **b.** Input and output voltages measured under periodic actuation of an HG-A sample. **c.** The output voltage as a function of hydrogel stiffness and thickness. The COMSOL-simulated sound pressure level intensity (a.u.) in the hydrogel is shown in the inset for HG-A. **d-e.** Water extraction rates from HG-A and HG-C gels and the corresponding energy consumption during ultrasonic actuation with 70% duty cycle. **f.** Anomalous diffusion model describes the moisture extraction process under ultrasonic actuation. **g-h.** Extraction rates and the associated energy costs during actuation with 30% duty cycle. **i.** Cyclic stability of the desorption behavior of HG-A hydrogel. Scale bar: 16 mm.

We used a hydrophone to measure the transmittance and attenuation of the ultrasound wave emitted from the actuator (see Supplementary Fig. S23) and passing through the hydrogel[55] (**Fig. 5a**). **Figure 5b** illustrates the signal captured by the hydrophone while it is in contact with the hydrogel when the actuator is driven periodically, ON for 10 seconds, OFF for 5 seconds (see also Supplementary Movie 6). These measurements revealed increased acoustic wave attenuation in thicker hydrogel samples (as expected[59]) and for hydrogels with higher storage moduli (**Fig. 5c**). The HG-C hydrogel, with 100 times higher crosslinker content than HG-A, exhibits a much higher storage modulus and about the same loss modulus ($G_A' = 1098, G_A'' = 240;\ G_C' = 2855, G_C'' = 241$), translating into higher attenuation of the ultrasound wave (**Fig. 5c**). HG-M hydrogel demonstrated a similar trend of the output voltage decrease with the increased thickness and stiffness (Supplementary Fig. S24). The COMSOL-simulated 2D image of the actuator-hydrogel system is shown in the inset for HG-A. It shows the high-intensity sound pressure level at the hydrogel-actuator interface, which attenuates away from the actuator surface. These observations are in agreement with prior studies of medical-grade hydrogels[60].

Next, we compared the extraction efficiency of HG-A and HG-C samples with comparable weight (0.2g and 0.19g, respectively) and equal moisture content equilibrated at ~80% RH (**Figs. 5d-e**). HG-A exhibited a higher weight loss of ~22% and energy consumption of ~12 MJ/kg under the 10-minute-long stimulation by the PZ-C actuator with ~114 Vpp and 70% duty cycle, while HG-C lost only ~12% of its weight and exhibited a much higher energy consumption of 21 MJ/kg under the same conditions. We fit the experimental data by using the Korsmeyer-Peppas (K-P) kinetic model, which establishes an exponential dependence of the fractional mass release on time $t$ (see Methods): $m(t)/m_0 = ke^{nt}$ [61,62] (**Fig. 5f**). This model is commonly employed to analyze the release mechanism of pharmaceutical drugs from polymers, when more than one release phenomena are involved ($k$ is a constant incorporating the sample structural and geometric characteristics, and the value of release exponent $n$ quantifies the release process). The fitted values of the release exponent $n$ were $0.65 \pm 0.1$ for HG-A $0.82 \pm 0.1$ for HG-C, indicating a swelling-controlled diffusion mechanism, rather than the Fickian (static) diffusion with $n = 0.45$[63].

The non-Fickian (dynamic) behavior observed in our experimental data can be characterized as an anomalous diffusion, and indicates the presence of an additional molecular relaxation process under a combined action of ultrasound, mechanical pressure, and heat[64]. A similar anomalous diffusion is observed for the ultrasound-triggered release of antibiotics from hydrogel carriers[65]. The ANOVA tests indicate that there is no statistically significant variance ($p > \alpha = 0.05$) between the $n$-values of the K-P model fitting for HG-A and HG-C samples (Supplementary Fig. S25). However, a higher fitted $n$ value for HG-C suggests that, owing to its high modulus, the water release mechanism in this hydrogel deviates further from the Fickian behavior than that in HG-A, translating into higher energy consumption. Supplementary Fig. S26 reveals the differences in the evolution of vibrational modes associated with both water molecules and polymer functional groups like methylene ($-CH_2-$) and amino group ($-NH_2$) during the sorption-desorption cycle, which underlie the differences in the relaxation process dynamics of hydrogels of varying stiffness.

To quantify synergistic and separate impacts of Joule heating and mechanical actuation on moisture extraction, we measured the moisture extraction rate in a cycle of periodic activation and deactivation of the PZ-C piezo-actuator (**Figs. 5g,h**). The plots reveal a persistent reduction of AWH-H weight even during the actuator off time periods, which indicates a temperature-driven water vapor release due to the high vapor pressure of the water in the hydrogel[51]. With this hybrid approach, we further increased extraction efficiency and reduced the energy consumption (to 10.2 MJ/kg and 12.4 MJ/kg for HG-A and HG-B, respectively). These data also reveal that increased mechanical rigidity of hydrogels reduces the rate of water release not only during the ultrasonic actuation but also by evaporation. The thermogravimetric analysis confirmed that when exposed to the same heating under the same environmental conditions, the stiffer HG-B and HG-C specimens released less moisture than HG-A (Supplementary Fig. S27). Despite the differences in the extraction efficiency, both HG-A and HG-C hydrogel samples exhibited consistent absorption and desorption cycles, indicating cyclic stability (**Fig. 5i** and Supplementary Fig. S28), while the intermediate-stiffness HG-B samples exhibited unstable performance (Supplementary Fig. S29) and were excluded from further analysis.

Comparative analysis of the data in **Fig. 4** and **Figs. 5d-i** suggests that the water extraction process can be further improved by the duty cycle and sample size optimization. We used our complete assembled system shown in **Figs. 1c-d** (see also Supplementary Fig. S30 and Supplementary Movie 7) to probe extraction efficiency from HG-A and HG-C specimens of varying moisture content. HG-A specimens were equilibrated at 77% RH, and then actuated for 25 minutes with a duty cycle of ~30-32%, equivalent to 8.75 minutes of piezo actuation time. **Figure 6a** reveals the reduction of energy consumption down to 5.25 MJ/kg for a medium-heavy (~1.73 g) sample, which covers the entire surface of the membrane. Similarly, for HG-C hydrogel, the optimized energy consumption of 6.2 MJ/kg was achieved with a ~1.65 g specimen (**Fig. 6b**). Note that while operating at ~33% duty cycle reduced the energy consumption, it also increased the actuation process duration (with the same amount of water extracted during 10.5 minutes at 70% duty cycle and 25 minutes at ~30% duty cycle, see Supplementary Fig. S31 and Supplementary Movie 7). Accordingly, the duty cycle needs to be optimized for environmental conditions, to compromise between the water harvesting amount per day and the corresponding energy consumption per unit mass of water.

To further optimize the system performance, we (i) reduced the actuation time to 2 minutes, (ii) performed a test using three identical actuators and samples simultaneously under 2-minute extraction cycles, and (iii) tested a single large HG-A sample with a diameter of 34 mm under multiple consecutive 2-minute extraction cycles. Under these optimized conditions, during each 2-minute-long extraction cycle in each of the tests, energy consumption was recorded to be below the thermal limit of 2.25 MJ/kg[66], thus yielding extraction efficiency in excess of 100% (see Supplementary Note 5 for detail). The lowest energy consumption of 0.48 MJ/kg has been recorded for the extraction from the largest 34-mm-diameter sample, for which cyclic sorption-desorption testing over 10 hours has been performed, with ten 1-hour-long sorption cycles run at 75% RH, separated by eleven 2-minute-long extraction cycles. **Fig. 6c** shows the extracted water mass and energy efficiency for each extraction cycle.

**Figure 6d** summarizes the extraction efficiencies recorded under different actuating cycles and with different sorbent samples described above and compares these values to the highest-efficiency prior literature data. Recent thermodynamic models of thermally-driven atmospheric water extraction predict the highest realistic thermal efficiency of 60%, corresponding to thermal energy consumption to 3.75 MJ/kg[45]. Real-world devices demonstrate even lower efficiencies, with the highest recorded efficiency of 10%[67], and the currently achieved minimum energy consumption of 22.5 MJ/kg. In stark contrast, our system exhibits energy consumption as low as 0.48 MJ/kg, corresponding to the efficiency of 428.6% (**Fig. 6c**), because it is not fundamentally restricted by the thermal limit inherent to evaporation-driven techniques.

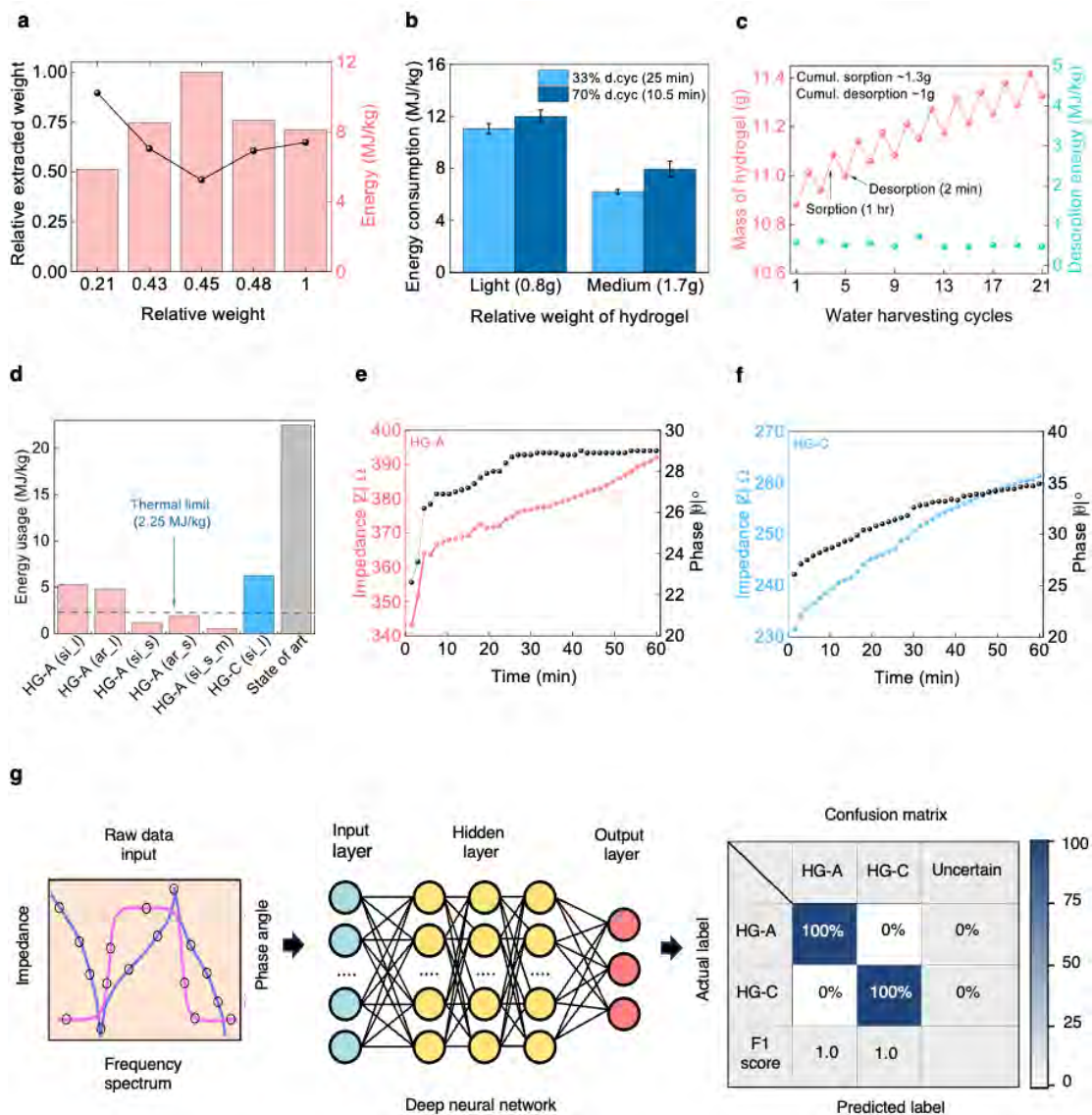

Figure 6. Strategies for the performance optimization, price of water produced, and in-situ self-correction. a-b. Minimization of the energy consumption by optimizing the sample mass (**a, b**) and the actuator duty cycle (**b**). **c.** Water sorption and desorption cycles of a large HG-A sorbent (34 mm diameter) at 25% duty cycle. **d.** A comparative analysis of energy consumption between actuator-enabled water extraction and the state-of-the-art thermally driven evaporation process[42]. Here, si_l = a single device actuated over a long period; ar_l = an array device actuated over a long period; si_s = a single device actuated over a short period; ar_s = an array device actuated over a short period; si_s_m = a single device actuated over multiple short periods. **e-f.** Dynamic changes of HG-A (**e**) and HG-C (**f**) samples monitored *in-situ* through impedance-phase angle spectroscopy. **g.** A schematic illustration of the machine learning model training on the measured impedance and phase angle data. The model uses a deep neural network for classification, which is shown in the confusion matrix.

## Discussion and outlook

Given the observed dependence of the extraction energy efficiency on the sample moisture content, *in-situ* monitoring and dynamic optimization of the sorption-extraction cycle under variable environmental conditions could help to improve the system performance during the field operation. Fourier-transform infrared spectroscopy (FTIR) may be used to monitor the changes in the sorbent properties (Supplementary Note 3, Supplementary Fig. S26, and Supplementary Table 3), but it cannot be deployed for continuous *in-situ* monitoring. Fortunately, the piezoelectric actuator itself can be used not only for moisture extraction but also for real-time process monitoring. Piezoelectric materials exhibit both direct and inverse piezoelectric effects, translating into their ability to either convert applied force into measurable electric quantity or convert electric potential into quantifiable displacement, respectively[68]. By taking advantage of the direct piezoelectric effect, the actuator can be used as an *in-situ* sensor[69,70]. This sensing capability is illustrated in **Figs. 6e-f**, which show changes in the impedance (|Z| Ω) and phase angle (|θ| °) of a PZ-C actuator loaded with either HG-A or HG-C hydrogel during the process of atmospheric moisture adsorption at ~20% RH and 20°C.

To evaluate the potential of using this sensing modality for continuous *in-situ* monitoring and control of the sorption process, we developed a deep learning model and demonstrated its capability to recognize absorption patterns of different hydrogels (**Fig. 6g**). Impedance variations datasets were collected during atmospheric moisture absorption cycles at ~20% RH and 20°C. We used 77% datasets for training and the rest 23% for testing. Following 200 epochs of training, the model successfully classified the impedance data and was able to distinguish between absorption patterns of HG-A and HG-C hydrogels, achieving an impressive prediction accuracy of 100% as shown in the confusion matrix (**Fig. 6g**). This monitoring and classification capability shows promise in designing fully automated systems that can switch from sorption to extraction once the optimum conditions are reached. Combining embedded sensing capability with deep learning tools for real-time data classification will allow for identifying abnormalities in the system performance and adjusting the system operation to the changes in the environment.

In our devices, we estimate the energy consumption to be roughly equally divided between the mechanical actuation (49%) and heating (51%), see Supplementary Note 9. The parasitic heat generated by a piezo actuator is used to heat a sorbent, and still contributes to the water extraction, but with lower efficiency than mechanical actuation. The effective electromechanical coupling coefficients of our prototype actuators was quite low, ranging from 17% to 19%. By designing a 1-3 composite piezoelectric transducer array device instead of a single ring-shaped actuator, an electromechanical coupling coefficient of 35% can be achieved[71]. High-efficiency single crystal piezoelectric materials exhibit even significantly higher electromechanical coupling factors, which can be complemented by the optimized transducer array design, promising further efficiency enhancement.

This study demonstrated moisture extraction by vibrational actuation with high efficiency from two different hydrogel systems. These results give reason to expect that ultrasound actuation can facilitate moisture extraction from other sorbents. This hypothesis is independently supported by literature evidence on increased drying rate of textiles[72], seaweed[73], and zeolite[74] under ultrasonic actuation. Each material system would need to be evaluated separately under optimized actuation conditions, moisture content, and sorbent geometry, which will be a subject of our future work. Under the new extraction scenario, the kinetics of both the sorption and desorption processes under the optimized vibrational actuation for each type of sorbent will play an important role. Accordingly, the known sorbents, including other types of hydrogels, metal organic frameworks, micro- and nano-fiber mats, and combinations thereof will need to be re-evaluated and re-engineered to improve their compatibility with the new extraction method. Internal nano- and micro-scale structure of sorbents will further play an important role in the efficiency of this process and need to be investigated (e.g., size and shape of the pores, surface-to-volume ratio, stiffness, etc). The overall shape of the sorbent sample can be engineered to provide acoustic resonances matched to the actuator driving frequencies. Finally, each sorbent may exhibit its overall highest efficiency when actuated not only by a different frequency but also around a different point in its sorption curve, calling for further fundamental studies and optimization effort.

We emphasize that ultrasonic water atomizers used for humidification and medical applications apply ultrasound to atomize bulk water or extract drugs from over-saturated medical hydrogels with very high free water content. The novelty of our work is in the application of piezo-induced ultrasonic actuation technology for water extraction from AWH sorbents, which have low moisture content, making efficient extraction much more challenging. Our work offers the first ever (to the best of our knowledge) experimental demonstration of the energy consumption of such process below the thermal limit. This major improvement in energy efficiency may promote AWH technology commercialization.

Our testing confirms that hydrogels maintain structural integrity (see Supplementary Note 6) after multiple actuation cycles. Measured Li ion concentration in extracted water varied between 1.5 and 15.2 ppm (or mg/L), comparable to the concentrations measured a recent study of bottled water in multiple countries (see Supplementary Note 7). While there are no regulatory thresholds for lithium either in the USA or globally, our future work will focus on the synthesis and testing of Li-free sorbents with high water uptake to limit the risk of water contamination and to reduce the cost of the technology. The presence of Li ions plays no role in the energy efficiency of water extraction, and our work demonstrates that our methodology is equally applicable to Li-free hydrogels such as HG-M.

We estimated the daily yield of water that can be harvested from a scaled-up system with a 1-m$^2$ sorbent area [L/m²/day] by extrapolating the results of cyclic sorption-desorption testing of a single prototype device with a large sorbent over 10 hours at 75% RH, i.e., by using the data shown in **Fig. 6c**. This simple scaling strategy assumes that multiple identical actuators with hydrogel samples identical to those used in our study are deployed in the form of a 2D array with the total area of the sample surface equal to 1 m$^2$. We further assumed that each actuator is powered separately and the energy consumption per device is the same as measured in our experiments. The resulting average daily productivity value has been predicted to be ~3.2 [L/m²/day], and the average device extraction efficiency - 0.576 MJ/kg (see Supplementary Note 5).

This is likely not the most efficient way of scaling the system up and overestimates the daily energy consumption while underestimating the energy efficiency of extraction. The effectiveness of large-area PZT-based ultrasound transduction has been demonstrated in various unrelated studies. For example, an interconnected network comprising hundreds of miniaturized PZT actuators has been employed over a 20cm x 20cm area, utilizing only 6W of electric power, resulting in efficient transduction intensity at a depth up to 8mm[75]. Similarly, larger piezoelectric components could be utilized in arrays, akin to those employed in sonar applications for underwater wireless communication, which are inherently designed with a larger footprint area extending up to several sq. feet[76]. Future research could implement these strategies to test the AWH sorbents with a custom-made actuator array over a large surface area and associated electronic circuits for optimum power supply to the array, which may yield higher energy efficiency than we predict based on the simple approach described above.

Finally, our preliminary technoeconomic analysis of the scaled-up system performance indicates the viability of commercializing our system, with the estimated cost of water of 0.19 $/L, which is lower than the cost of bottled water in various countries worldwide, including USA, UK, Canada and Australia (see Supplementary Notes 7-9 and Supplementary Fig. S41).

## Acknowledgements

This work has been supported by the Abdul Latif Jameel Water and Food Systems (J-WAFS) Lab at the Massachusetts Institute of Technology (MIT). I.I.S. thanks his mother (late Jahanara Begum) and extends gratitude to Waleed Akbar, Tammy Lee, Joseph Paradiso, Mitchell Thompson, Steven Nagle and Mark Feldmeier for useful discussions, and Rohit Karnik for access to his laboratory facilities. This work was carried out in part by using MIT.nano and ISN facilities at MIT.

## Data availability

The data supporting the plots within the main text are available in Supplementary Information. Additional data relevant to this study are available from the authors upon request.

## Contributions

S.V.B. conceived the idea and supervised the research. C.D.D-M. and I.S. prepared the hydrogel samples. I.S. performed all the material and device characterizations. I.S. and C.L. designed, prototyped, and tested the portable collection system. M.C. and M.L. characterized and analyzed the vibrational modes of actuators. I.S. carried out numerical modeling. I.S. and C.D.D-M. performed techno-economic analysis. I.S., C.D.D-M., and S.V.B analyzed and interpreted the results. I.S. drafted the manuscript with input from all other authors. S.V.B. and C.D.D-M. reviewed and edited the manuscript.

## Competing interests

A U.S. Provisional Patent Application has been filed (inventors: S.V. Boriskina and I. Shuvo).

## Materials and Methods

### Hydrogel synthesis

We synthesized polyacrylamide hydrogels based on a simple, one-pot approach, where polymer, salt, initiator, crosslinker, and accelerator, were all mixed in water. For all hydrogels, we first dissolved the 16.72 g lithium chloride (≥99%) salt in a beaker with 50 mL deionized (DI) water. The solution was continuously mixed with a magnetic stirring bar, covered to prevent evaporation, and then left to cool down to room temperature. We added sequentially 4.18 g of acrylamide (≥99%), the N,N'-methylenebisacrylamide (MBA) (99%) as crosslinker (50 mg for HG-C, 2.5 mg for HG-B and 0.5 mg for the HG-A) and 14.2 mg of ammonium persulfate (APS) (≥98%) as initiator. We degassed the mixed solution for 10 minutes under vacuum in a dessicator. Finally, we added 12 µL of N,N,N',N'-tetramethylethylendiamin (TEMED) (≥99%) as accelerator. 4 mL of solution were poured into a petri dish, covered with the lid, and left to gelate at room temperature overnight. All the chemicals were purchased from Sigma-Aldrich and used as received.

### Sorption evaluation

The dynamic vapor sorption isotherms were characterized using an environmental simulation chamber (BINDER GmbH). Initially, the samples were dried at a temperature of 95 °C for a duration of 24 hours using a gravity convection oven (MTI Corporation). Subsequently, the weight of the samples that had been dried in the oven was measured using an analytical balance (OHAUS Corporation). Next, the oven-dried samples were treated inside the environmental chamber under varying RHs (15%, 30%, 55%, 85%) at a constant 25°C temperature and at a maximum stage time of 720 min, ensuring that the specimen weight reached an equilibrium state. In addition to our designed hydrogel, we also used a commercially available hydrogel (HG-M) (a cross-linked glycerol- 2-Acrylamido-2-methylpropanesulfonic acid sodium salt hydrogel, Medela). Finally, the water uptake was quantified by calculating the ratio of the water content in the sample at equilibrium to its oven-dried weight. The findings are shown as the mean ± standard deviation.

### Oscillatory shear rheology characterization

Rheological experiments were conducted using a rheometer (HR-20, TA Instruments). For all the experiments, a 25mm parallel plate geometry was employed to ensure accurate loading on the hydrogel specimens. A circular mold was used to fabricate gelled discs, which were formed into a diameter of 25 mm for utilization underneath the parallel plate. Before conducting the tests, the samples were given a five-minute period to reach a state of equilibrium, ensuring that both mechanical and thermal factors were balanced. The temperature was controlled using a built-in Peltier system. The experiments involving the oscillatory frequency and strain/amplitude sweeps were carried out at 25°C. Temperature sweep experiments were conducted at an angular frequency of 6.28 rad/s and 0.1% strain.

### Thermal analysis

Thermogravimetric analysis was conducted to evaluate the desorption behavior and thermal stability of the hydrogels using a thermal analysis system (TGA 5500, TA Instruments).

Approximately 10 mg of samples were put into a titanium pan in the presence of nitrogen gas and were heated gradually at a rate of 1°C/min until they reached a temperature of 50°C.

## Surface morphology

The surface morphology of the hydrogel specimens was examined using a scanning electron microscope (SEM) (Zeiss Sigma 300 VP). Hydrogels were completely dehydrated to prevent off-gassing during sputtering and SEM imaging. To prepare the samples and visualize them under the SEM, they were stuck to aluminum stubs with double-sided carbon tape. Next, a sputter coater (Desk V, Denton) was used to apply a 10 nm thin film coating of gold/palladium (Au/Pd) (60:40) in presence of argon gas.

## Forward-osmosis test for water permeability study

A conventional diffusion cell was used to conduct the forward osmosis test and assess the efficacy of water penetration through the porous stainless-steel membranes employed in this investigation. The experimental configuration closely resembled a recent study[77]. The porous membrane was installed between two glass cells, each having a capacity of 7ml. The left cell is referred as the NaCl-solution cell, and the right one is the DI-water solution cell, (see Supplementary Fig. S8). A 2M NaCl solution was made by dissolving 0.81g of NaCl in 7ml of DI water. The right cell contained 7 ml of DI-water with conductivity in the µS (micro-siemens) range. As the water conductivity increased, reaching the mS (milli-siemens) range, the ion mass began to transport through the membrane from the ion-cell to the DI-cell. This occurred under vigorous mixing at a speed of 1350 RPM at room temperature, driven by the concentration gradient. The conductance measurements were conducted using a conductivity probe (SevenCompact, Mettler Toledo) for a duration of 10 minutes for each of the three SS membranes.

## Piezoelectric ultrasonic device architecture

The solid housing for the piezoelectric transducer was modelled in SOLIDWORKS and parts were printed using a Bambu Lab X1-Carbon Combo 3D printer. Aside from the glass domes, all parts within the assembly have a hole with a 12 mm diameter in the center to allow droplets to fall into the bottom dome since water can be harvested from the hydrogel on both sides (top and bottom) of the piezoelectric transducer. Two-ring structures, top and bottom insulators, are used to hold two glass domes in place. Cuts are made into the assembly to allow the domes to slide into the case. The top insulator is 8 mm tall, has an outer diameter of 33 mm, and an inner diameter of 25mm, allowing for a perfect fit of the glass dome, which had an outer diameter of slightly below 25mm. The wire track that houses the piezoelectric device is 8 mm tall and has an outer diameter of 20mm and an inner diameter of 17 mm. Also, it has a cut that is 2mm deep to hold the piezoelectric ring and provides a cavity to allow wiring and ambient moisture access into the case. The support platform connects the top and bottom insulators. The bottom insulator and the tripod together host the bottom glass dome. Each leg of the tripod is 38mm long.

The piezoelectric transducer used in our system was circular in shape (see Supplementary Table 1-2 for detail), where a piezoelectric ring made of lead zirconate titanate (PZT) polycrystal was affixed to a porous stainless-steel membrane. To fabricate the transducer for our system, the membrane and the PZT rings were purchased from Dongguan Norvis Electronic Corporation and securely fastened. The PZT was affixed to the porous membrane by clamping one side, while the other side was covered with a hydrophobic epoxy resin. Porous membranes with three different nozzle sizes (10, 70, and 100 µm) were used with PZT crystals of two different frequencies to build the low (~110 kHz) and high (~165 kHz) frequency actuators. Piezo drivers were used to actuate the transducer at two different duty cycles of 70±3% and 30±3%, operating at RMS voltages of 40 and 30 V, respectively. The driver boards were powered at 1.5 W using a DC power supply to drive the piezoelectric transducers.

## Piezoelectric charge coefficient d$_{33}$ measurement of the piezoelectric materials

A Berlincourt meter was employed to measure the piezoelectric coefficients, d$_{33}$ (pC/N), of the PZT crystals used in fabricating the four transducers. During the measurement of the coefficients, both a static force and a dynamic force are exerted on the sample. The tool controls and maintains a constant dynamic force of 0.25N/110Hz, while the static force of around ~1N was manually controlled. The static force is measured using a static force sensor to assure consistent measurements (see Supplementary Fig. S32 for experimental setup).

## Electrical impedance, resonant spectra, and electromechanical coefficients of piezo transducers

The electrical phase-impedance spectrum of the transducer devices was measured using an impedance spectroscopy (Solartron SI 1260, AMETEK). The impedance spectrum was screened to obtain two key parameters, the anti-resonance (f$_a$) and resonance (f$_r$) frequencies. The evaluation of the effectiveness of piezoelectric devices in converting electric energy into mechanical energy was conducted based on their effective electromechanical coupling coefficient (k$_{eff}$)[56,71] using the following equation: $k_{eff} = \sqrt{(f_a^2 - f_r^2)/f_a^2}$.

## Modal, deflection, and velocity analysis of ultrasound devices

The vibrational modes of the steel membrane and PZT ring were analyzed using a PICOSCALE Vibrometer (SmarAct Metrology GmbH & Co. KG), which is a laser scanning vibrometer employing confocal optics and Michelson interferometry. The sample was actuated using the built-in signal generator of the vibrometer. In this case, the device was connected to a general-purpose input/output (GPIO1 from the vibrometer controller), which is used to output an electrical signal and directly feed it to the device using an adapted BNC cable.

First, microscopic imaging was used to identify the specific area of interest on the membrane and PZT ring for the modal analysis (see Supplementary Fig. S33). The membrane was analyzed at two distinct locations: the core and the periphery. Measurements on the periphery of the membrane and PZT were consistently conducted at a significant distance from the electrode cable. To uncover the local resonances, a linear sweep excitation ranging from 1 Hz to 500 kHz, with a duration of 0.01 s and an amplitude of 1 Vpp was generated and output via the GPIO1. For each identified local resonance, a 2D modal analysis was performed. During this analysis, the interferometric laser beam is raster-scanned over the sample while the amplitude and phase of the vibrations are extracted using dual-phase lock-in demodulation.

For this application, a single-frequency sinusoid is output at GPIO1, which is also used as the reference signal for the lock-in amplifier. The recorded time series is post-processed using the FFT (Fast Fourier Transform) algorithm to convert the signal from time domain into frequency domain in the form $x(t) = A \sin(\omega t)$ to reduce the noise and reveal the resonances. The velocity ($v$) of the device at its resonance frequency ($f_d$) was then calculated as $v = dx/dt = \omega \cdot A \cdot \cos(\omega t)$, where $x$ is the displacement signal and $\omega$ is the angular frequency. Maximum velocity value is then estimated as $v_{max} = \omega \cdot A = 2\pi f_d A$. The amplitude $A$ of the signal can be easily retrieved by performing the Fourier analysis of the recorded time-series, and the RMS value of the velocity at the peak amplitude value is calculated as $v_{RMS} = 0.7 v_{max}$.

In the case of a linear sweep excitation, the energy of the signal is spread over a large bandwidth. In order to retrieve the real amplitude of the motion, a correction factor is applied: $k = \sqrt{(f_1 - f_0)dt}$, where $f_1$ and $f_0$ are the highest and the lowest frequencies, respectively, and $dt$ is the duration of the sweep. In our calculations, a correction factor of $k = 70.71$ was used. During the experimental analysis and 2D imaging of vibrational modes, the displacement signal is demodulated on-the-fly using a digital dual-phase lock-in amplifier, and no correction factor to retrieve the actual amplitude is needed. The resulting 2D image of the sample deflection is post-processed to retrieve the maximum displacement value.

### Elemental mapping, surface roughness imaging, and atomization photography

Elemental mapping, surface roughness imaging, and atomization photography Zeiss Sigma 300 VP was used to capture SEM images of the membrane nozzles and PZT material. The tool also employed Energy Dispersive X-ray (EDX) technology for elemental mapping of the PZT devices and provides precise information about their composition. A laser scanning confocal microscope (VK-X250, Keyence) was used to do non-contact surface profiling and roughness assessment of the membrane after the extraction cycle. The Nikon D5300 DSLR camera was used to capture the atomization process of water as it was expelled from the hydrogel and funneled through the nozzles of the porous membrane incorporated into the transducer. A green crossline was generated using a 520 nm green laser module (OXLasers), with the focal point positioned below the nozzle axis. This improved the process of envisioning for photography. T-slotted aluminum extrusions were used to build a platform for mounting the transducer between two L-shaped joint brackets to perform imaging experiments.

### Infrared imaging

To monitor and visualize hydrogel heating caused by the PZT transducer heat generation, an infrared (IR) thermal camera (FLIR ETS320, Teledyne FLIR LLC) was utilized. IR imaging was also used to monitor the process of hydrogel cooling to ambient temperature after the actuation process. Due to the challenge of IR-imaging a highly polished surface such as a metallic stainless-steel membrane in our device, we used Dr. Scholl's™ spray to deposit a thick opaque layer onto the membrane, and during imaging, emissivity (ε) value of 0.95 was set in the IR camera.

## Ultrasonic radiation sensitivity

We used an acoustic sensor hydrophone (H Instruments) to measure and quantify the force exerted by a piezoelectric transducer as a function of voltage (V). To conduct the acoustic experiment, the transducer was placed within a petri dish and the hydrogel samples were attached to the device. The acoustic sensor was linked to its amplifier board operating at 9V that had a gain of 25 dB with bandwidth up to 120kHz and placed on the hydrogel surface. The transmitted and received waveforms were recorded. Our device served as the transmitter, while the acoustic sensor functioned as the receiver. The resulting output voltage was measured for hydrogel samples of varying thickness.

## Comparative weight-loss study between Joule heater membrane and the transducer

We used a multi-channel calibrated commercial K-type thermocouple (HT-9815, RISE PRO) to measure the temperature of a free-standing (i.e., not clamped to the PZT) stainless steel membrane used as the Joule heater. The transducer device received a power supply of 1.5 W in order to maintain a consistent temperature for a comparative analysis. Throughout the experiment, either Joule heating or ultrasonic actuation were applied for 1 minute each, followed by a subsequent short deactivation period to weigh the amount of weight loss. The weight loss (%) of the hydrogel was calculated by comparing the extracted mass to the initial mass of the sample.

## Vibrational spectra of wetted hydrogels

Fourier transform infrared (FTIR) spectroscopy was used to investigate the changes in the vibrational spectra of hydrogels with different level of cross-linking and varying moisture content. FTIR with a built-in attenuated total internal reflectance (ATR) capability was used (iS50, ThermoFisher). The raw FITR spectra were deconvoluted into distinct Gaussian peaks corresponding to individual vibrational modes of polymer and trapped water molecules.

## Energy consumption and efficiency calculation

The energy consumption, E [MJ/kg], required for moisture extraction was calculated as follows: $E = P/m = ((\text{power}, \; kWh) \cdot (duty \; cycle, \%) \cdot 3.6 \cdot 10^6 \cdot 10^{-6})/10^{-3} \; [MJ \cdot kg^{-1}]$, where $P$ is input energy [MJ], and $m$ [kg] is the extracted mass. Using the energy consumption (E), the efficiency of the water release process of our device, $\eta$, is calculated as $\eta = h_{lv}/E$, where $h_{lv} = 2.257 \; [\text{MJ/kg}]$ is the enthalpy of vaporization of water (see Supplementary Note 4).

## Li content test of extracted water

Agilent 5100 DVD was used to test the quality of the ultrasonically-extracted water from PAM-LiCl hydrogels via a coupled plasma optical emission spectrometer (ICP-OES) elemental analysis, similar to the test methods reported in recent studies of AWH hygroscopic gels[30]. Prior to testing with the Agilent 5100 DVD, the extracted water was collected using VWR sterile syringe filters. Multiple desorbed specimens were tested.

### 3D electromechanical finite element modeling (FEM)

COMSOL Multiphysics® was used to simulate vibrational modes of the transducer device consisting of a PZT ring clamped to a steel membrane with nozzles. Solid Mechanics and Electrostatics modules were used to study the deformation dynamics, and a 3D asymmetrical model was constructed to facilitate visualization (see also Supplementary Note 2). Using a voltage sweep, the displacement and velocity were studied, and the phase-impedance spectrum was analyzed employing a frequency sweep. Simulating the nozzle dispersion pattern akin to the manufactured membrane proved challenging in COMSOL. Consequently, all the nozzles were uniformly distributed and a diameter of 50 µm was used to accommodate around 800 nozzles. The simulation results for impedance, velocity, and displacement exhibited a high degree of agreement with the experimental data, without significant deviation. This simulation technique was also employed to visualize the behavior of ultrasonic radiation in hydrogel. For this study purpose, the Young's modulus (E) was determined from the storage modulus of our hydrogel utilizing the formula $E = 2G(1 + \mu)$, where $\mu$ represents the Poisson ratio and storage modulus was used as the magnitude for $G$. The density was estimated at approximately 1200 kg/m$^3$ according to literature[78], and a Poisson's ratio of 0.5 is often employed for hydrogels. Following the calculation, the inputs were entered into COMSOL to simulate the sound pressure induced by the actuator in the hydrogel volume.

### Deep neural network for the extraction dynamics monitoring

We used Edge Impulse machine learning platform[79] to build a deep neural network, which was used to categorize and distinguish between different types of hydrogels during the sorption process. The neural network was trained using the measured dynamic changes in the phase angle and impedance of the transducer loaded with hydrogel samples, which were continuously sorbing moisture at ~20% RH and ~19°C. The trained model has three hidden layers, with a total of 100 neurons, and exhibited accuracy of 100%. The training phase used 77% of the datasets, while the remaining 23% were allocated for testing the model.

Supplementary Materials for

# High-efficiency atmospheric water harvesting enabled by ultrasonic extraction

Ikra Shuvo[1], Carlos D. Díaz-Marín[2], Marvin Christen[4], Michael Lherbette[4], Christopher Liem[3], Svetlana V. Boriskina[2,*]

*[1]Media Lab, Massachusetts Institute of Technology, Cambridge, MA 02139, USA*
*[2]Department of Mechanical Engineering, Massachusetts Institute of Technology, Cambridge, MA 02139, USA*
*[3]Department of Electrical Engineering and Computer Science, Massachusetts Institute of Technology, Cambridge, MA 02139, USA*
*[4]SmarAct Metrology GmbH & Co. KG, Oldenburg, HRA 207391, Germany.*
**Corresponding author: sborisk@mit.edu**

1. Supplementary Notes

2. Supplementary Tables

3. Supplementary Figures

4. Supplementary Info- Movies

# 1. Supplementary Notes

## 1.1. Supplementary Note 1 - Piezoelectricity and constitutive piezoelectric equations

In our study, we utilized the <u>converse piezoelectric effect</u> (i.e., mechanical deformations of the material caused by the applied electric field) of an ultrasound transducer to operate it as an actuator to remove water from a hydrogel. We also proposed to utilize a <u>direct piezoelectric effect</u> (i.e., electrical charge generation under mechanical pressure) as a mechanism of in-situ monitoring and control of the moisture harvesting and extraction processes. Supplementary Fig. S1 schematically illustrates the piezoelectric and converse piezoelectric phenomena[1].

The fundamental interactions between the electrical and mechanical characteristics of piezoelectric materials and the surrounding environment are characterized by two constitutive piezoelectric equations, commonly known as the actuator and sensor equations[1,2] (Eq. S1):

$$\begin{bmatrix} S \\ D \end{bmatrix} = \begin{bmatrix} s^E & d \\ d & \varepsilon \end{bmatrix} \begin{bmatrix} T \\ E \end{bmatrix}. \tag{S1}$$

Actuator equation:

$$S = du/dx = s^E T + dE. \tag{S2}$$

Sensor equation:

$$D = dT + \varepsilon E. \tag{S3}$$

Here,

$D$ is the dielectric (charge density) displacement
$S = du/dx$ is the induced strain
$E$ is the electric field
$s^E$ is an elastic compliance at constant electric field
$T$ is stress
$E$ is the dielectric permittivity at constant temperature
$d$ is a piezoelectric strain constant

## 1.2 Supplementary Note 2 Numerical analysis using multi-physics modelling technique

PZT polycrystal was used in this study to make the actuating devices. It is a widely used commercial piezoelectric materials due to its high energy conversion rate and piezoelectric coefficients[1]. The strain-charge relationship for a PZT material could be expressed as[3]:

$$
\begin{bmatrix} S_1 \\ S_2 \\ S_3 \\ S_4 \\ S_5 \\ S_6 \end{bmatrix} = \begin{bmatrix} s_{11}^E & s_{12}^E & s_{13}^E & 0 & 0 & 0 \\ s_{21}^E & s_{22}^E & s_{23}^E & 0 & 0 & 0 \\ s_{31}^E & s_{32}^E & s_{33}^E & 0 & 0 & 0 \\ 0 & 0 & 0 & s_{44}^E & 0 & 0 \\ 0 & 0 & 0 & 0 & s_{55}^E & 0 \\ 0 & 0 & 0 & 0 & 0 & s_{66}^E = 2(s_{11}^E - s_{12}^E) \end{bmatrix} \begin{bmatrix} T_1 \\ T_2 \\ T_3 \\ T_4 \\ T_5 \\ T_6 \end{bmatrix} + \begin{bmatrix} 0 & 0 & d_{31} \\ 0 & 0 & d_{32} \\ 0 & 0 & d_{33} \\ 0 & d_{24} & 0 \\ d_{15} & 0 & 0 \\ 0 & 0 & 0 \end{bmatrix} \begin{bmatrix} E_1 \\ E_2 \\ E_3 \end{bmatrix} \quad \text{(S4)}
$$

$$
\begin{bmatrix} D_1 \\ D_2 \\ D_3 \end{bmatrix} = \begin{bmatrix} 0 & 0 & 0 & 0 & d_{15} & 0 \\ 0 & 0 & 0 & d_{24} & 0 & 0 \\ d_{31} & d_{32} & d_{33} & 0 & 0 & 0 \end{bmatrix} \begin{bmatrix} T_1 \\ T_2 \\ T_3 \\ T_4 \\ T_5 \\ T_6 \end{bmatrix} + \begin{bmatrix} \varepsilon_{11} & 0 & 0 \\ 0 & \varepsilon_{22} & 0 \\ 0 & 0 & \varepsilon_{33} \end{bmatrix} \begin{bmatrix} E_1 \\ E_2 \\ E_3 \end{bmatrix}. \quad \text{(S5)}
$$

Piezoelectric strain coefficient matrix for the PZT polycrystal has the following form[4]:

$$
[d_{ij}] = \begin{bmatrix} 0 & 0 & d_{31} \\ 0 & 0 & d_{32} \\ 0 & 0 & d_{33} \\ 0 & d_{24} & 0 \\ d_{15} & 0 & 0 \\ 0 & 0 & 0 \end{bmatrix}. \quad \text{(S6)}
$$

The governing equations of piezoelectricity used to calculate the vibrational modes of the actuating membrane and the acoustic wave propagation through hydrogel samples include the momentum equation (Newton's second law, Eq. S7) and the charge conservation equation (Eq. S8):

$$
\rho \frac{d^2 u}{dt^2} = \nabla \cdot S + F, \quad \text{(S7)}
$$

$$
\nabla \cdot D = \rho_V, \quad \text{(S8)}
$$

where $\rho_V$ is a volume charge density, $\rho$ is a solid density, $u$ is a solid displacement vector, $F$ if the force per unit volume.

The electric field is computed from the electric potential $V$:

$$
E = -\nabla V. \quad \text{(S9)}
$$

COMSOL Multiphysics (6.1) was used to calculate vibrational eigenfrequencies and the associated eigenmodes[5] of a piezo actuator driven by sinusoidal voltage.

## 1.3. Supplementary Note 3 – FTIR analysis of hydrogels

We used the Fourier-transform infrared spectroscopy (FTIR) to reveal the changes in the hydrogen properties underlying ultrasonic-driven water extraction in process (see Methods). PAM hydrogels contains major functional groups like methylene (-CH$_2$-), carbonyl (C=O) and amino group (-NH$_2$), which form hydrogen (H-) bonds with H$_2$O and may play a crucial role in the configuration of adsorbed water molecules[6]. The interplay of heat and kinetic pressure induced by ultrasonic actuation breaks some of these hydrogen bonds, facilitating water extraction. The resulting reduced moisture content is manifested by the shifts and intensity changes of the peaks in the FTIR spectra of hydrogels[7,8].

Supplementary Fig. S26 compares the FTIR spectra of (i) as-prepared (pristine) hydrogels (HG-A and HG-C), (ii) hydrogels after 30 minutes of ultrasound actuation (US 30 min), (iii) hydrogels exposed to the environment with 25% RH for 30 minutes at 21$^{\circ}$C following US actuation (Abs. 30 min), and (iv) pure DI water. Two broad absorptance bands observed in the FTIR spectra of HG-A and HG-C hydrogels between 2800 cm$^{-1}$ and 3800 cm$^{-1}$ and between 1500 and 1800 cm$^{-1}$ in Supplementary Figs. S26a and S26e, respectively, have been deconvoluted into individual Gaussian peaks corresponding to different vibrational modes of the polymer functional groups and water molecules[7–9] (Supplementary Figs. S26b-d,f-h, see Methods). Several peaks have been fitted and assigned to different vibrational modes based on prior literature data (see Supplementary Table 2.3)[9].

These data reveal that the water content in the sorbent material, which is varied under the piezoelectric actuation and moisture adsorption process, is manifested in the changes in vibrational peak positions and relative intensities. A more detailed analysis of FTIR spectra of different sorbents under different harvesting and ultrasonic actuation scenarios is expected to offer valuable insights into the molecular dynamics of the polymer chains and their interactions with water molecules. Such an analysis will be a subject of a separate study focused on the development of optimized ultrasound-responsive water harvesting hydrogels with thermodynamics-limit-breaking energy efficiency moisture extraction performance.

## 1.4. Supplementary Note 4 – Efficiency calculation of the device

The efficiency of the water release process, $\eta$, was calculated as [10,11]:

$$\eta = \frac{h_{lv}}{E} \times 100\%$$

where $h_{lv} = 2.257 \, [\text{MJ/kg}]$ is the enthalpy of vaporization of water and $E \, [\text{MJ/kg}]$ is the specific energy consumption defined as the energy input (electrical, thermal, or a combination thereof, depending on each individual system configuration) divided by the mass of water released from a sorbent. In an ideal thermal process, all the heat provided to the system is used to evaporate water i.e., $E = h_{lv}$, where it is also considered that the enthalpy of desorption is identical to the enthalpy of vaporization of water. Therefore, an ideal thermal process has $\eta = 100\%$. In this work, we refer to this ideal thermal process with $\eta = 100\%$ as the thermal limit. In a real thermal water release process, the heat provided is used not only evaporate water, but also to sensibly heat the system and sorbent, and is partially lost to the ambient, leading to $\eta < 100\%$.

For our experiments, the energy consumption ($E$) after the weight-loss experiment was calculated using the following equation, where $P$ is input energy, $m$ is the output extracted weight or amount of weight loss measured in $g$, and $1\ Wh = 3.6 \cdot 10^3\ J$:

$$E\left[\frac{MJ}{kg}\right] = \frac{P}{m} = \frac{\text{Supplied power } [MJ]}{m\ [kg]}$$

$$= \frac{V[V] \cdot I[A] \cdot \text{Actuation time } [h] \cdot \text{Duty cycle} \cdot 3.6\left[\frac{J}{Wh}\right] \cdot 10^3}{m\ [g] \cdot 10^{-3}}$$

As an example, when we applied 5 V and 0.3 A at 70% duty cycle for 10 min (~0.17 hr) using our piezoelectric actuator to desorb water from HG-M (Fig. 4E), the weight loss was measured to be $0.13\pm0.002$ g. Then, the energy consumption per unit mass of water extracted can be calculated as:

$$E = \frac{5[V] \cdot 0.3[A] \cdot 0.17\ [h] \cdot 0.7 \cdot 3.6\left[\frac{J}{Wh}\right] \cdot 10^3}{0.13\ [g] \cdot 10^{-3}} = 4.84\left[\frac{MJ}{kg}\right] = 1.35\left[\frac{kWh}{kg}\right]$$

Then, the efficiency of the water release process, $\eta$, can be calculated as:

$$\eta = \frac{h_{lv}}{E} \times 100\% = \frac{2.25\left[\frac{MJ}{kg}\right]}{4.84\left[\frac{MJ}{kg}\right]} \times 100\% = {\sim}46.4\%$$

Note that this is the device-level energy efficiency, which explicitly captures all the energy losses during the extraction system operation. In the manuscript, we demonstrate that this efficiency depends on the actuation time and decreases during longer actuation periods. Cycling testing of the system under different actuation periods revealed that short actuation times (2 minutes) yield the highest extraction energy efficiency and even allow energy consumption lower than the thermal limit.

We compared our results with the specific energy consumption $E$ and moisture extraction energy efficiency $\eta$ data for several experimentally demonstrated atmospheric water harvesting systems[10]. These values are summarized in Table S4, which is shown below. The table also shows the estimates of the moisture extraction efficiency per unit mass per day. Note that, unlike the specific energy consumption $E$ and extracted water mass $m$, the daily values $E_d$ and $m_d$ are either measured over a daily collection/extraction cycle or estimated by accounting the time needed to collect the moisture from the atmosphere before it can be extracted. All the values in the table represent the experimentally obtained system-level performance data in a daily harvesting cycle and, as such, can be compared on equal footing.

Note that under the above standard definition of the efficiency ($\eta$) of AWH systems, an efficient non-thermal extraction method can have a specific energy consumption lower than the enthalpy of vaporization of water, $h_{lv}$. In this case, the energy efficiency of the water release process can exceed 100%. This is the case measured and reported in our manuscript, which we describe as 'breaking the thermal limit.

### 1.5. Supplementary Note 5 – Scaled-up AWH system performance estimate

To evaluate the daily water yield and energy efficiency of extraction for an AWH device scaled to the 1-m$^2$ area, we implemented the simplest scale-up strategy, which assumes that multiple identical actuators with hydrogel samples identical to those used in our study will be deployed in the form of a 2D array with the total area of the sample surface equal to 1 m$^2$. We further assume that each actuator will be powered separately and the energy consumption per device will be the same as measured in our experiments. This is likely not the most efficient way of scaling the system up and would overestimate the daily energy consumption of the system and underestimate the energy efficiency of extraction.

To illustrate the validity of the simple scaling-up approach, we first performed a test using three identical actuators simultaneously (Fig. S34). Three hydrogel samples weighing about 2.6g, after sorption at 75% relative humidity, were simultaneously stimulated by three piezoelectric ultrasonic devices operating at a 25% duty cycle for 10-min. In all cases, the specimens exhibited a loss above 0.1 g. The measured specific energy consumption ranged from 1.9 to 2.1 MJ/kg after 2-min and from 4.2 to 5.8 MJ/kg after 10-min actuation (Fig. S35), comparable to the same values measured in a single-actuator device shown in Table S4.

The daily yield of water from a 1-m$^2$ sorbent area [L/m$^2$/day] has been calculated by extrapolating the results of cyclic sorption-desorption testing of a single prototype device over 10 hours at 75% RH. The piezo actuator has a surface area of 0.0002 m2 (diameter = 16mm), and the sorbent has a surface area of 0.0009 m2 (diameter = 34 mm).

Fig. 6c shows the extracted water mass and energy efficiency for each extraction cycle (with ten 1-hour-long sorption cycles separated by eleven 2-minute-long extraction cycles). During each 2-minute-long extraction cycle, energy consumption was recorded to be below the enthalpy of water vaporization, thus yielding extraction efficiency in excess of 100%.

To estimate the water yield per device per day, we first calculated the average sorption and desorption rates by using the data shown in Fig. 6c and Fig. S36 (a). The average collection and extraction rate is around 0.13 g/hr and 0.08 g/2min (Fig. 6c and S36a). The extraction rate exceeds the collection rate by a factor of ~18.5, and during a 1-hr harvesting period, the sorbent collects more water than can be extracted in 2 min. These data indicate that to maximize and balance both the sorption and desorption processes over the course of one day, the system may be operated in 34 sorption-extraction cycles per 24 hours, each cycle comprised of about 40 minutes of collection followed by 2 minutes of extraction (see Table S5). Under this operational scenario, all the harvested moisture will be extracted with low energy expenditure, as it takes about 2 minutes to extract the whole amount of moisture harvested over 40 minutes at 75%RH. For operation at RH different than 75%, a different sorption-desorption cycle ratio may need to be implemented, to be optimized separately for each RH level.

To extrapolate the water yield to a 1-m$^2$ sorbent area, this number was further scaled up from one device having 0.0009-m$^2$ sorbent area to a 1-m$^2$-area array. The resulting average daily productivity value has been predicted to be ~3.2 [L/m$^2$/day]. Note that with our approach, harvesting can be done during both night and day (unlike the devices relying on solar-driven evaporation). In turn, the device efficiency of 0.16 kWh/kg (0.576 MJ/kg) was calculated by using the average mass extracted from each of two hydrogel specimens over eleven 2-min-long extraction cycles (see Table S5).

Fig. S36 (b) illustrates an alternative method of device operation, consisting of continuous harvesting for 24 hours at 75%RH, followed by desorption involving 25 consecutive cycles of 2-minute-long extractions over a total duration of 50 minutes (i.e., without a sorption period in between the cycles), resulting in the desorption of 1g of water at an energy cost of 1.13 MJ/kg. This energy expenditure is nearly a factor of two greater than the average energy cost observed for the 2-minute-long extraction cycles depicted in Fig. S36 (a).

These methods together demonstrate the adaptability of our system, which is not constrained by temporal limitations, unlike the solar-derived desorption mechanism that operates based on the availability of sufficient or proper daylight hours at a slower desorption rate in comparison to our system. We chose the lower-energy-expenditure approach illustrated in Fig. 6 (c) to estimate the daily yield and the cost of the harvested water.

We conducted similar studies with smaller specimens (≤16 mm diameter) and observed similar cyclic stability and energy efficiency as shown in Fig. S37. It can be seen that in this case the collection rate is much slower than the extraction rate, and during a 1-hr harvesting period, the sorbent collects less water than can be extracted in 2 minutes of actuation (see Table S6). As such, harvesting time imposes the upper limit on the daily water yield.

In this case, to balance the mass of harvested and extracted water over a 24-hour cycle, we assume 12 sorption-extraction cycles per 24 hours, each comprised of ~ 2 hours (118 min) of collection followed by ~2-minutes of extraction. During 118 min of collection, slightly less water is absorbed per device than can be extracted in 2 minutes, so we assume that each extraction period will be slightly less than 2 minutes long. The average daily water yield per device at 75%RH is calculated as the amount of water collected over twelve 118-min-long cycles. To extrapolate the water yield to a 1 $m^2$ sorbent area, this number was further scaled up from one device area to 1 $m^2$. The resulting average daily productivity value has been predicted to be ~1.5 [L/$m^2$/day], and the average energy consumption - 0.411 kWh/kg (1.48 MJ/kg).

## 1.6. Supplementary Note 6 – Structural integrity of hydrogel under ultrasonic extraction

To evaluate the structural integrity of the hydrogel and the extracted water quality after desorption, the water extracted from LiCl-PAM hydrogels (HG-A) were filtered using a VWR@ sterile syringe filter, commonly employed for ICP-OES testing. In this experiment, three piezoelectric devices were used to extract water from a ~5.92 g HG-A specimen at a duty cycle of 25-30% over a duration of ~45 min. The setup and test were made and conducted within an enclosed petri dish, respectively. Following desorption, ~0.21 ml of desorbed water was collected from the petri dish lid and subsequently filtered using a VWR filter paper kit to examine for any traces of polymer residue.

The initial dry weight of the filter paper kit was ~2.99 g, which increased to ~3.02 g after the filtration of ~0.21 ml extracted water at 20% relative humidity, resulting in wetting the filter paper. Following drying at room temperature, the weight of the filter paper returned to 2.99 g, with no polymer residue detected. This provides the evidence that hydrogels remained intact following the ultrasonic actuation process, and the extracted water is not contaminated by the polymer.

## 1.7. Supplementary Note 7 – Li content in the water extracted from LiCl-PAM hydrogels

We additionally analyzed Li content in the water extracted from LiCl-PAM hydrogels. There are no regulatory thresholds for lithium either in the USA or globally, and EPA doesn't have a specific health advisory for lithium in a drinking water [12,13]. While typical therapeutic oral doses of Li are as high as 600–1200 mg/day [14], Li may pose a potential concern for human health, and some studies recommend to limit casual Li-ion intake to 3.1 mg per day[15]. At the same time, recent studies of bottled water from dozens of countries revealed a wide range of Li content, from ~0 to 9.9 mg/L [13,16].

We conducted the inductively coupled plasma optical emission spectrometer (ICP-OES) elemental analysis on the water desorbed from a LiCl-PAM hydrogel to test its quality. The resultant Li-ion concentration varied between 1.5 and 15.2 ppm (or mg/L), see Supplementary Fig. S38. However, the presence of Li ions plays no role in the energy efficiency of water extraction, and our work demonstrates that our methodology is equally applicable to Li-free hydrogels such as HG-M. Our future work will focus on the synthesis and testing of Li-free sorbents with high water uptake to limit the risk of water contamination and to reduce the cost of the technology.

### 1.8. Supplementary Note 8 – Contributions of mechanical actuation and heating of the device

By modifying the technoeconomic feasibility study conducted by Zhong et al.[11] on sorption-based atmospheric water harvesting (SAWH) system, we carried out a preliminary technoeconomic analysis of our system. Contrary to that study, we consider that the lifespan of the material and device are distinct. In addition, we included the cost of supplied electric power into cost of water production (**Cost of water**, in units of \$/L), which is calculated as:

$$\text{Cost of water} = \text{ Capital expenditure (CAPEX)} + \text{ Operational expenditure (OPEX)}$$

$$= \frac{\left(\frac{(\text{Cost}_{\text{hydrogel}} \cdot \rho_{\text{hydrogel}} \cdot V_{\text{hydrogel}}}{\tau_{\text{hydrogel}}} + \frac{\text{Cost}_{\text{device}}}{\tau_{\text{device}}}\right)}{\text{Yield}} + (\text{Electricity cost} \cdot \text{Energy intensity})$$

$$(\text{Eqn. S11})$$

Here,

$\text{Cost}_{\text{hydrogel}}$ = cost of hydrogel-salt composite per kg of material in the dry state,

$\rho_{\text{hydrogel}}$ = density (2000 kg/m$^3$)[17] of the hydrogel-salt composite in the dry state,

$V_{\text{hydrogel}} = A_{\text{hydrogel}} \cdot H_{\text{hydrogel}}$ = volume of the hydrogel-salt composite in the dry state,

$A_{hydrogel} = 1$ m² area of the hydrogel-salt composite, making cost per m² of our overall system,

$H_{hydrogel}$ = typical hydrogel-salt composite thicknesses used in our device ~1mm (=0.001m),

$\tau_{hydrogel}$ = hydrogel lifetime in days (30-365 days), which is one of the variables of analysis,

$Cost_{device}$ = cost of the SAWH device excluding the sorbent cost,

$\tau_{device}$ = lifetime of the piezoelectric PZT actuator (20-45 years), which is another variable [18,19]

**Yield or productivity** = amount of water produced by day (L/m²/day),

**Electricity cost** = 0.17 $/kWh, electricity price based on US Bureau of Labor Statistics [20],

**Energy intensity** = supplied energy per liter of desorbed water, expressed in kWh/L

Table S7 summarizes all the costs of the hydrogel-salt composite ingredients and an actuator unit as well as the energy cost of operating the actuator. With these data, we estimate the $Cost_{hydrogel}$ and $Cost_{device}$ are 0.4-1.1 $/kg and 1111-2222 $/m² (based on a single device used for a gel with 34 mm diameter), respectively.

The daily yield of water of 3.2 kg/m²/day has been estimated from the results of the cyclic sorption/extraction experiment as detailed in Supplementary note 5 above. CAPEX, OPEX, and cost of water were then estimated to be 0.05-0.09 $/L, 0.1 $/L, and 0.19 $/L (Fig. S39) assuming the device lifespan of 25 years and hydrogel lifespan of 30 days (Fig. S40). Based on a worldwide survey conducted on 2023, the cost of a fresh bottled water, ranges between 0.23 and 1.35 $/L around the globe as shown in Fig. S40, and thereby, estimated to have an average price of 0.79 $/L (data taken from [21]).

## 1.9. Supplementary Note 9 – Contributions of mechanical actuation and heating of the device

With the hardware configuration shown in the circuit diagram in Fig. S41 (a), we measured that approximately 300 mW of the supplied 1.5W DC power enters the piezoelectric ultrasound device. Based on our estimated effective coupling coefficient for the PZT device, the actuator efficiency ($\eta_a$) — that is, the ratio of electrical energy consumed to the mechanical energy produced — is $\eta_a = k_{eff}^2 = 3.6\%$. In the case of perfect coupling (i.e., $\eta_a = 100\%$), the mechanical loss tangent, $\tan\varphi$, would have been $1/Q = 1/32 = 3.1\%$ (where Q is the quality factor of PZT-5H ceramic), and 3.1% of the input power would be lost as heat[22,23]. Since the actual coupling efficiency is only 3.6%, the actual heat losses are 3.1% of 3.6% plus another portion of the 3.6% associated with work done by the PZT. We assume that the combination of the work done on the hydrogel by the PZT system during its contraction and expansion and the internal mechanical losses of the PZT material itself comprise roughly 25% of the mechanical energy, which is lost as heat, i.e., 25% of 3.6% (i.e., 0.9% of 300 = 2.7 mW). The remaining energy is attributed to mechanical storage, calculated as 2.7% (=3.6% - 0.9%) of 300 mW, which is equal to 8.1 mW.

As piezoelectric devices are essentially capacitors, the bulk of the delivered electrical energy is stored energy, which ultimately returns to the power source. The power delivered to the piezoelectric capacitor is about 96.4% (=100% - 3.6%) of the 300 mW input power, which is 291 mW.  The dielectric loss tangent, $\tan\delta_e$, for PZT-5H piezoelectric ceramics is about 2%, and the lost power is dissipated as heat, generating about 5.8mW, i.e., 2% of 291 mW.  Hence, the combined mechanical and dielectric power loss to heat is therefore about $5.8\ mW + 2.7\ mW = 8.5\ mW$ when the actuator is driven at its resonant frequency.

In summary, we estimate the energy consumption to be roughly equally divided between the mechanical actuation and heating, with individual contributions measuring 8.1 mW (49%) and 8.5 mW (51%), respectively, while 283.4 mW were returned to the power source (Fig. S41b). Due to the complexity of quantifying the dynamic interaction between the ceramic, the metallic membrane, and the hydrogel during actuation we offer this relatively simplistic analysis to quantify the individual contributions of work, mechanical losses, and dielectric losses.  Future work will aim to develop a more comprehensive model to decouple the different effects of heat and vibration on different actuators and sorbents.

## 2. Supplementary Tables

2.1. Supplementary Table 1. Geometry of the metal membranes and piezoelectric rings.

| Device | Frequency | Piezoelectric ring | | | Membrane mesh | | | |
|--------|-----------|---------------------|---------------------|---------------------|------------------|--------------------------------------|-----------|------------------|
| | (KHz) | Thickness (mm) | Outer diameter (mm) | Inner diameter (mm) | Thickness (mm) | Membrane & mesh diameters (mm) | Nozzle no | Nozzle size (μm) |
| PZ-A | 107-113 | 0.6 | 16 | 7.8 | 0.05 | 16, 5 | 1000 | 10 |
| PZ-B | 107-113 | 0.6 | 16 | 7.8 | 0.05 | 16, 5 | 1000 | 70 |
| PZ-C | 107-113 | 0.6 | 16 | 7.8 | 0.05 | 16, 5 | 1000 | 100 |
| PZ-D | 165-170 | 0.6 | 11.6 | 5 | 0.05 | 13.8, 5 | 1000 | 10 |

2.2. Supplementary Table 2. Mass of the actuator and its major components.

| Device | PZT ring (g) | Membrane (g) | Entire actuator (g) |
|--------|--------------|--------------|---------------------|
| PZ-A | 0.715±0.013 | 0.079±0.012 | 1.172 ±0.019 |
| PZ-B | 0.709±0.011 | 0.078±0.008 | 1.169 ±0.016 |
| PZ-C | 0.713±0.011 | 0.077±0.005 | 1.169 ±0.011 |
| PZ-D | 0.381±0.012 | 0.059±0.004 | 1.088±0.012 |

## 2.3. Supplementary Table 3. Vibrational mode assignments in the FTIR spectra of HG-A hydrogels.

| Peak | Wavenumber | Description |
|------|------------|-------------|
| 1 | 3584 $cm^{-1}$ | Asymmetric O–H stretching modes of weakly or non-hydrogen-bonded water molecules (HBs of the water molecules are partially or entirely broken) |
| 2 | 3504 $cm^{-1}$ | Symmetric O–H stretching modes of weakly or non-hydrogen-bonded water molecules (HBs of the water molecules are partially or entirely broken) |
| 3 | 3427 $cm^{-1}$ | Out-of-phase O–H stretching modes of water molecules with four HBs |
| 4 | 3353 $cm^{-1}$ | Asymmetric stretching modes of $NH_2$ |
| 5 | 3279 $cm^{-1}$ | In-phase O–H stretching modes of water molecules with four HBs |
| 6 | 3206 $cm^{-1}$ | Symmetric stretching modes of $NH_2$ |
| 7 | 3129 $cm^{-1}$ | An overtone of the O–H stretching mode at 3261 $cm^{-1}$ |
| 8 | 3106 $cm^{-1}$ | C–H stretching modes of the polymer chains (three modes, i.e., one C–H stretching mode of the CH groups and symmetric and asymmetric stretching modes of the $CH_2$ groups) |

2.4. Supplementary Table 4. A summary of variables derived from literature to derive specific energy consumption and the efficiency of the devices for water release process in comparison to our devices.

| Reference | Water production (kg/day) | Energy input (kWh/day) | Specific energy consumption (kWh/kg) | Efficiency $\eta$ (%) |
|---|---|---|---|---|
| LaPotin et al. [24] | 0.06 | 0.421 | 7.02 | 9.5 |
| Hanikel et al. [25] | 0.41 | 10.79 | 26.32 | 2.4 |
| Xu et al. [26] | 0.023 | 0.36 | 15.65 | 4.0 |
| Shan et al. [27] | 0.311 | 3.475 | 11.17 | 5.6 |
| Almassad et al. [28] | 0.72 | 12.456 | 17.3 | 3.6 |
| This work for HG-M [a] | 0.00013* | 0.000175* | 1.35 | 47 |
| This work for HG-A [b] | 0.00013* | 0.000188* | 1.44 | 43 |
| This work for HG-A [c] | 0.00014* | 0.000188* | 1.44 | 43 |
| This work for HG-A [d] | 0.00004* | 0.000013* | 0.31 | 200 |
| This work for HG-A [e] | 0.00007* | 0.000038* | 0.54 | 117 |
| This work for HG-A [f] | 0.00094 | 0.000138 | 0.15 | 428.6 |

* Per cycle

[a] After 10-min actuation cycle using a single device

[b] After 25-min actuation cycle using a single device

[c] After 10-min actuation cycles using an array made of three devices

[d] After 2-min actuation cycle using a single device

[e] After 2-min actuation cycle using an array made of three devices

[f] Measured over 11-hr period utilizing a large specimen and eleven 2-min actuation cycles (i.e., 22 min over 11hr)

## 2.5. Supplementary Table 5. A summary of water harvesting cycles of larger HG-A (>16 mm diameter) sorbent.

|  | HG-A Specimen 1 | HG-A Specimen 2 | Average values |
|---|---|---|---|
| Collection per device per hour [g] | 0.12 | 0.13 | 0.125 |
| Collection per device per 40 minutes [g] | 0.08 | 0.087 | 0.0835 |
| Collection per device for 34 cycles [g] | 2.77 | 2.96 | 2.865 |
| Extraction per device per 2 min [g] | 0.074 | 0.085 | 0.0795 |
| Extraction per device for 34 cycles [g] | 2.52 | 2.91 | 2.715 |
| Collection per $m^2$ per day for 34 cycles [kg] | 3.07 | 3.26 | 3.165 |
| Extraction per $m^2$ per day for 34 cycles [kg] | 2.77 | 3.19 | 2.98 |
| Specific energy consumption [kWh/kg] | 0.171 | 0.148 | 0.160 |

## 2.6. Supplementary Table 6. A summary of water harvesting cycles of smaller HG-A (≤16 mm diameter) sorbent.

|  | HG-A Specimen 1 | HG-A Specimen 2 | Average values |
|---|---|---|---|
| Collection per device per hour [g] | 0.012 | 0.013 | 0.0127 |
| Collection per device per 118 minutes [g] | 0.024 | 0.026 | 0.025 |
| Extraction per device per 2 minutes [g] | 0.029 | 0.033 | 0.031 |
| Collection per device per day [g] | 0.289 | 0.31 | 0.3 |
| Collection per $m^2$ per day [kg] | 1.447 | 1.549 | 1.523 |
| Specific energy consumption [kWh/kg] | 0.426 | 0.396 | 0.411 |

## 2.7. Supplementary Table 7. Parameters used in estimating the cost of water production.

| Component | Amount used | Price | Ref. |
|---|---|---|---|
| Lithium chloride * | 0.8 kg | 0.1-1 $/kg | [29] |
| Acrylamide | 0.2 kg | 1.6 $/kg | [30] |
| Ammonium persulfate | 0.68 g | 0.61 $/kg | [31] |
| N,N'-methylenebisacrylamide | 2.4 g | 1 $/kg | [32] |
| N,N,N',N'-tetramethylethylendiamin | 0.57 mL or 0.44 g [33] | 10 $/kg | [34] |
| Ultrasonic system * | 1 unit | 1-2 $/unit | [35] |
| Electricity cost ** | 0.00025 kWh | 0.0002 $/10min | [36] |

* The price fluctuated during the study duration

** Estimated for a single unit

## 3. Supplementary Figures

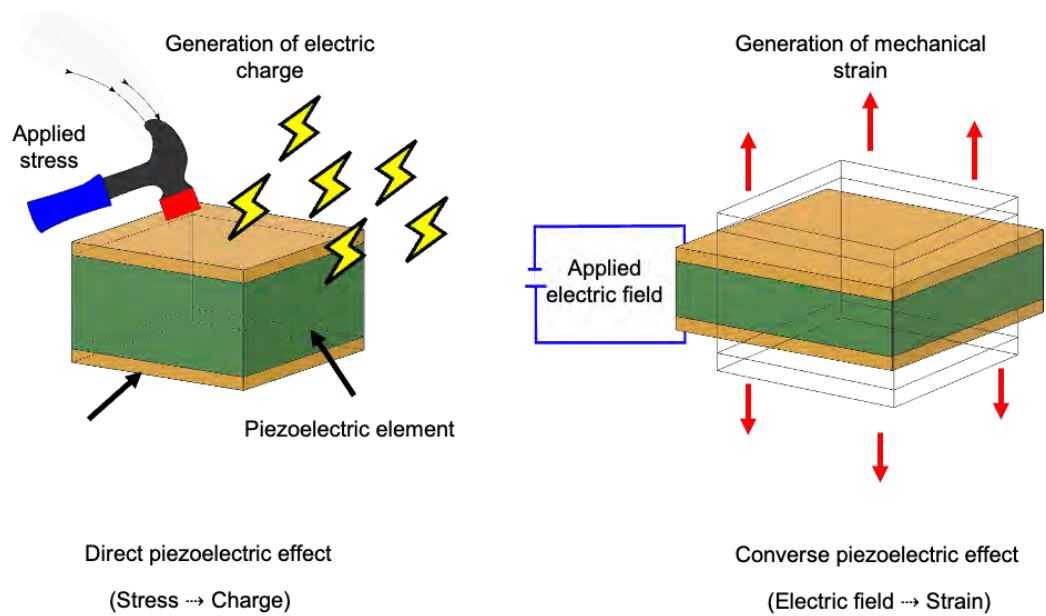

**Supplementary Figure S1. Direct (left-sensor) and converse (right-actuator) piezoelectric effects.** We utilized the converse piezoelectric effect to actuate the water-harvesting hydrogel placed on top of a piezoelectric transducer and to expel water out of it. In addition, we made use of the direct piezoelectric effect to achieve *in-situ* monitoring of the changes in the weight of the hydrogel throughout the process of atmospheric water harvesting. Subsequently, we employed the acquired sensing information to train our deep neural network designed for hydrogel classification.

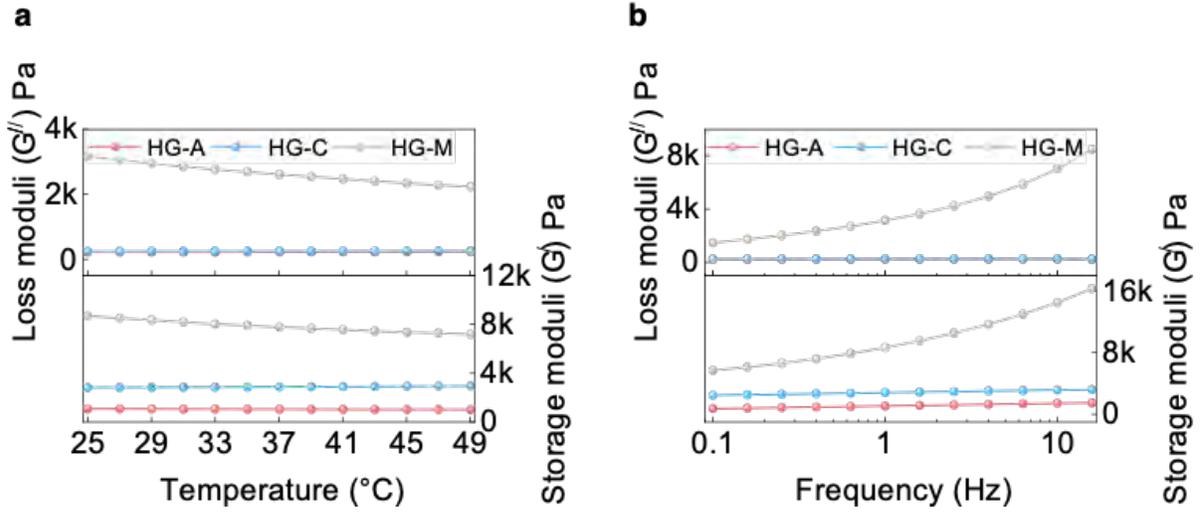

**Supplementary Figure S2. Rheological performance of the PAM-LiCl (HG-A and HG-C) and commercial hydrogels (HG-M). a-b**. Modulus changes of the hydrogel specimens as a function of temperature (a) and frequency (b).

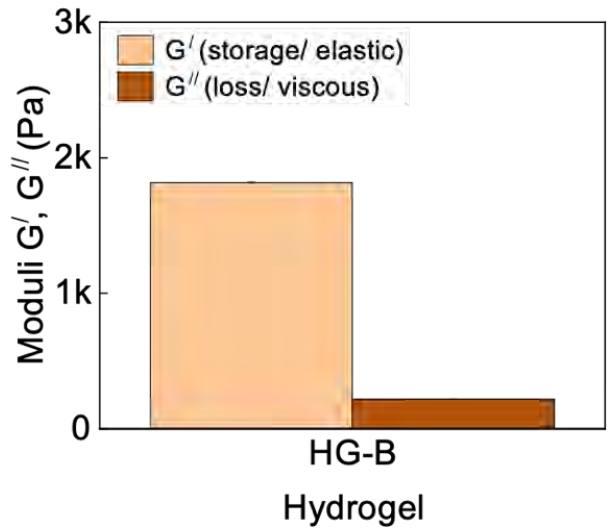

Supplementary Figure S3. Storage (light brown) and loss (dark brown) moduli of the HG-B hydrogel specimen.

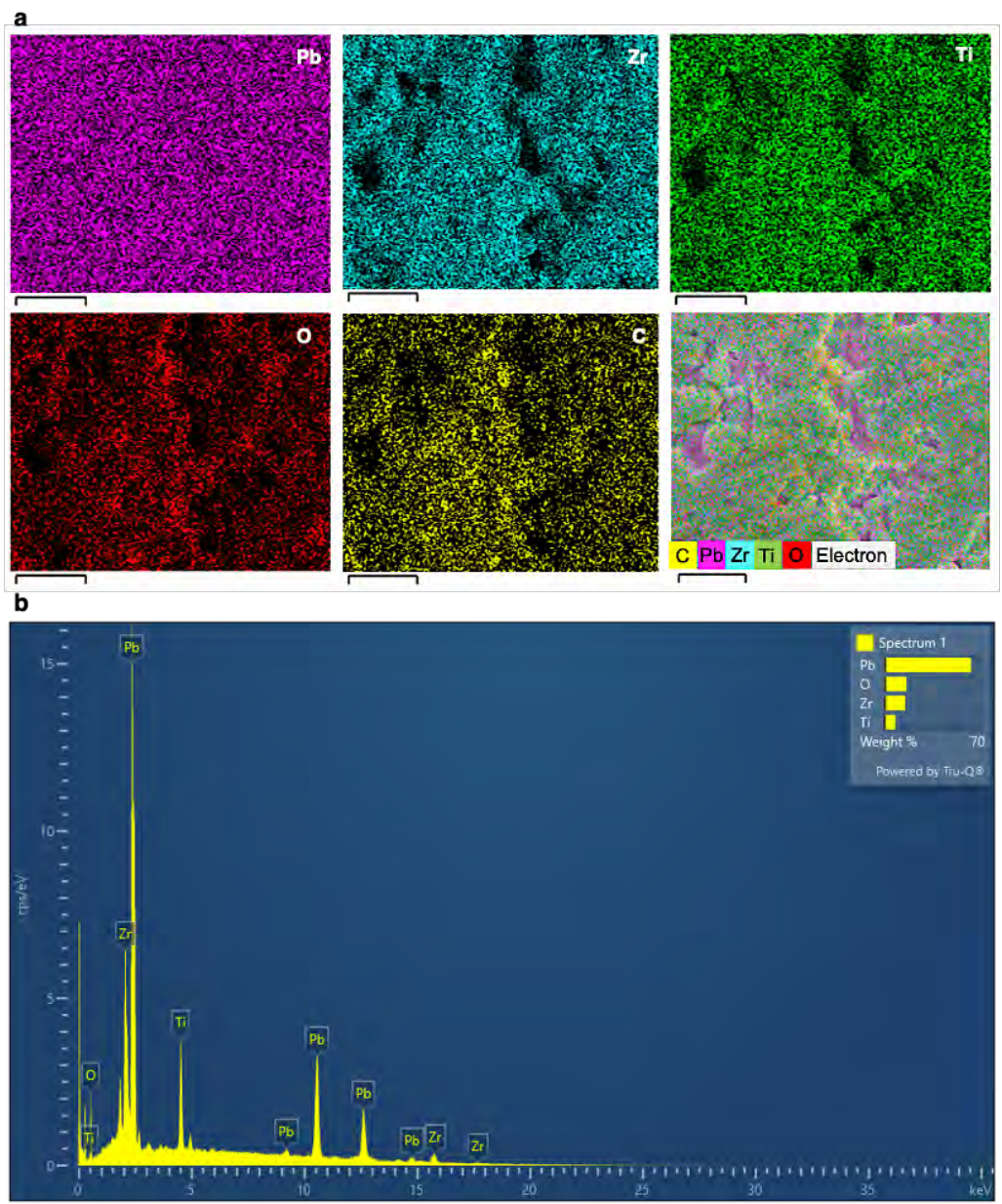

**Supplementary Figure S4. Energy dispersive X-ray spectroscopy (EDS) of PZT polycrystal used in this study**. **a**. Composition mapping of different elements (Pb, Zr, Ti, O, C) and their layered EDS image. **b**. The corresponding EDS spectra. Scale bars: 10 μm.

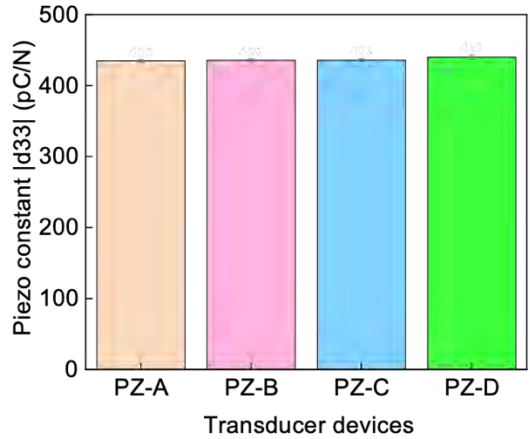

**Supplementary Figure S5. Measured piezoelectric charge coefficients (pC/N) of the bare ceramic materials employed in fabricating the four transducers (PZs- A, B, C, and D).** This comparative analysis demonstrates that despite minor variances in the operating frequency and geometry, the piezoelectric coefficients are comparable among all the transducers.

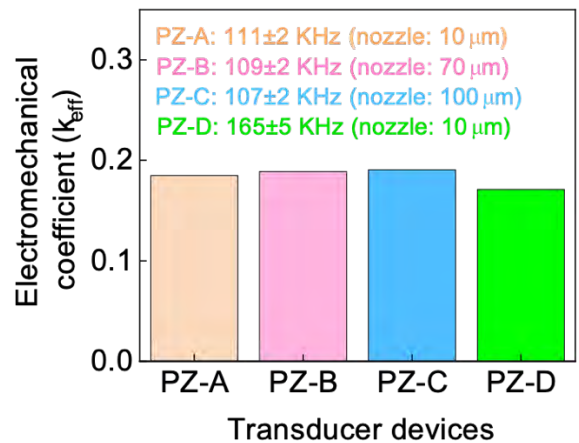

**Supplementary Figure S6. Effective electromechanical coupling co-efficient of the four actuator devices (PZs- A, B, C, and D).** This comparative analysis demonstrates that regardless of the nozzle size, the coupling coefficients are comparable for all the low-frequency transducers or actuators but differ from those of the high-frequency actuators.

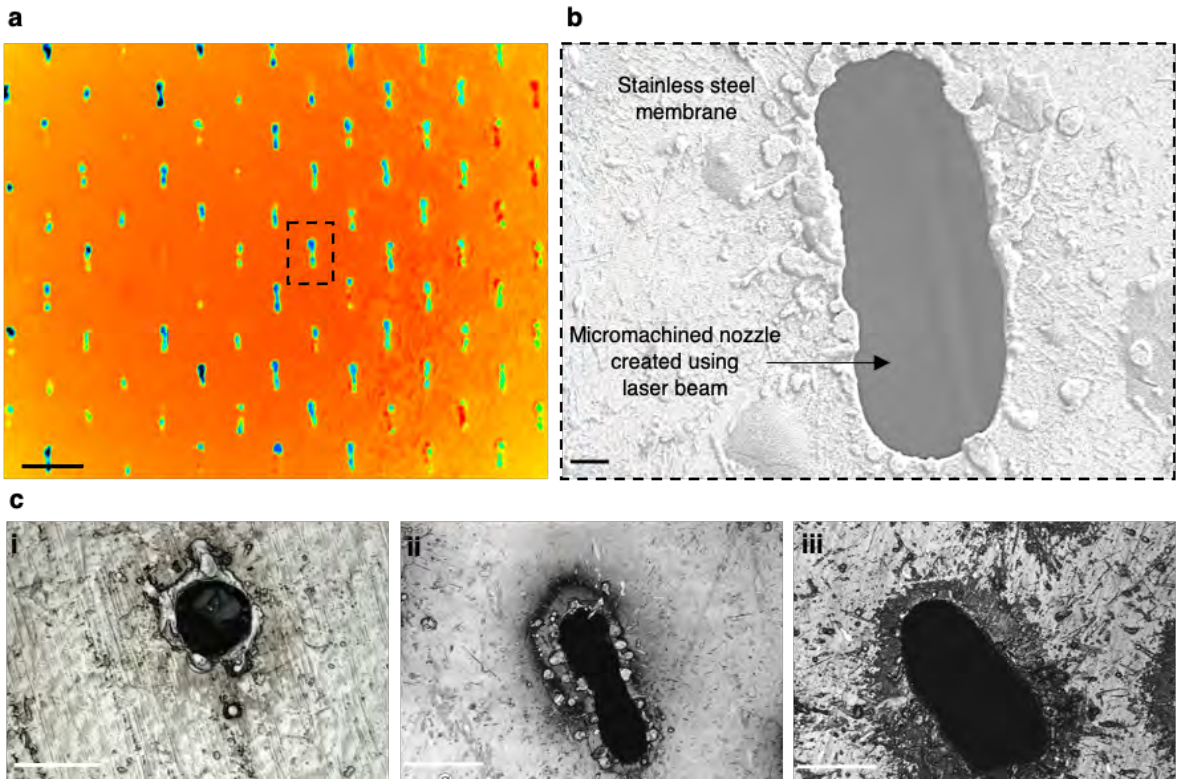

**Supplementary Figure S7. Microscopic imaging of porous membranes. a-b.** A heat map of a porous membrane surface (a) and a SEM image of a single nozzle of the membrane (b). **c.** Confocal laser scanning microscopy (CLSM) images of the three different nozzles used in three different membranes. Scale bars: 200 μm (a); 10 μm (b); 10 μm (c i); 30 μm (c ii); 50 μm (c iii).

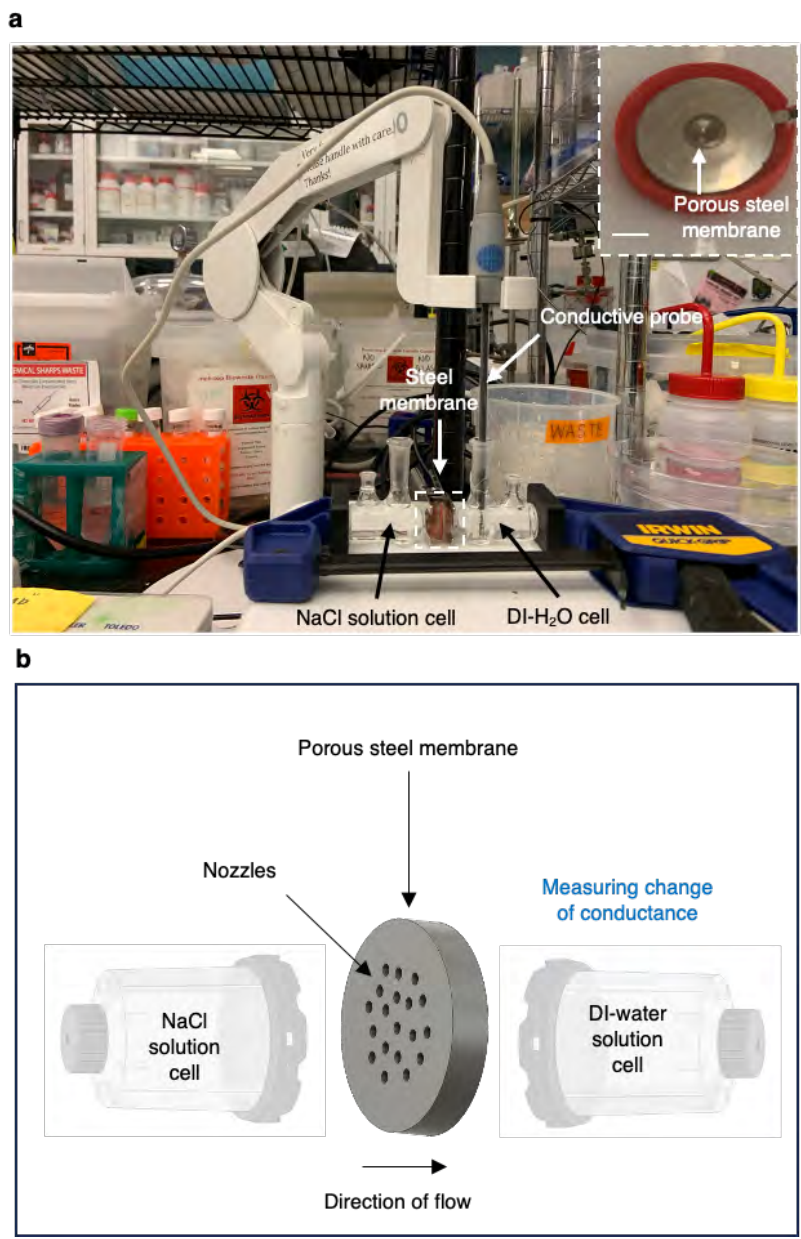

**Supplementary Figure S8. Ion-solution permeability test. a**. A photograph of a set-up for evaluating water transmission performance through porous membranes with different nozzle sizes. Inset shows an optical image of a bare porous stainless-steel membrane used in this study. **b**. Representative schematic illustrating the flow of ions from left cell to right cell causing change of conductance. Scale bars: 4 mm (a-inset)

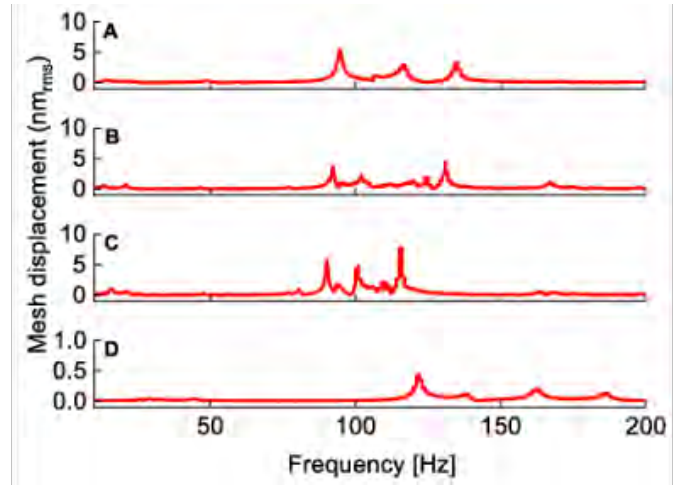

Supplementary Figure S9. Out-of-plane local (single point) modal analysis at the center of the membrane mesh (M) for PZs- A, B, C, and D transducers. The frequencies are obtained by performing a Fast Fourier transform (FFT) on the time-domain data.

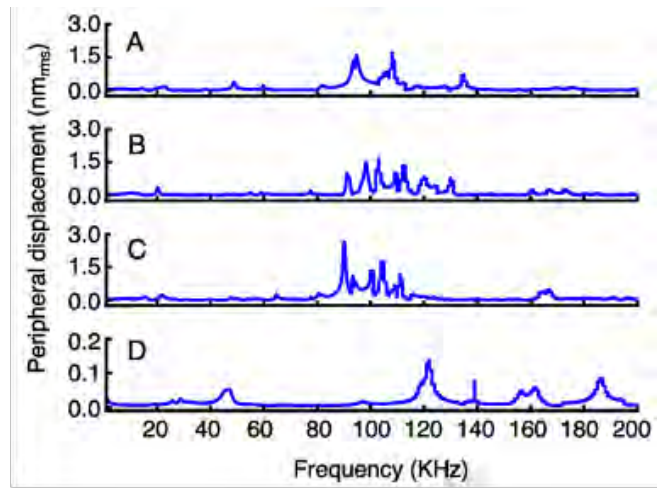

Supplementary Figure S10. Out-of-plane local (single point) modal analysis at the periphery (P) of the membrane for PZs- A, B, C, and D transducers. The resultant frequencies are obtained by performing a Fast Fourier transform (FFT) on the time-domain data.

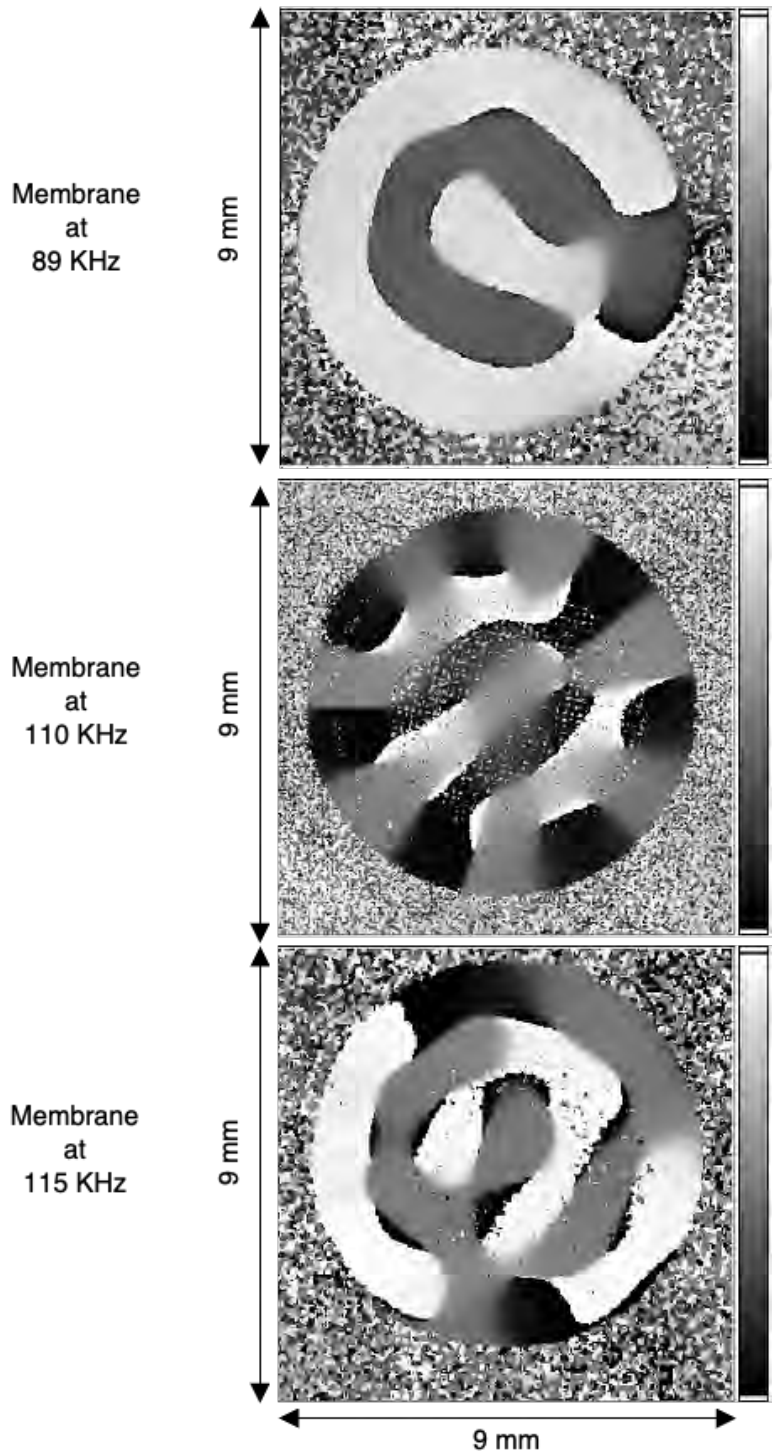

Membrane at 89 KHz

Membrane at 110 KHz

Membrane at 115 KHz

**Supplementary Figure S11. Out-of-plane phase change during the global modal analysis of the membrane.** The modal analysis was conducted at three different frequencies: 89, 110, and 115 kHz (from top to bottom). The corresponding representative animations of the out-of-plane deflection patterns of the vibrating membrane are included as Supplementary Movie 4.

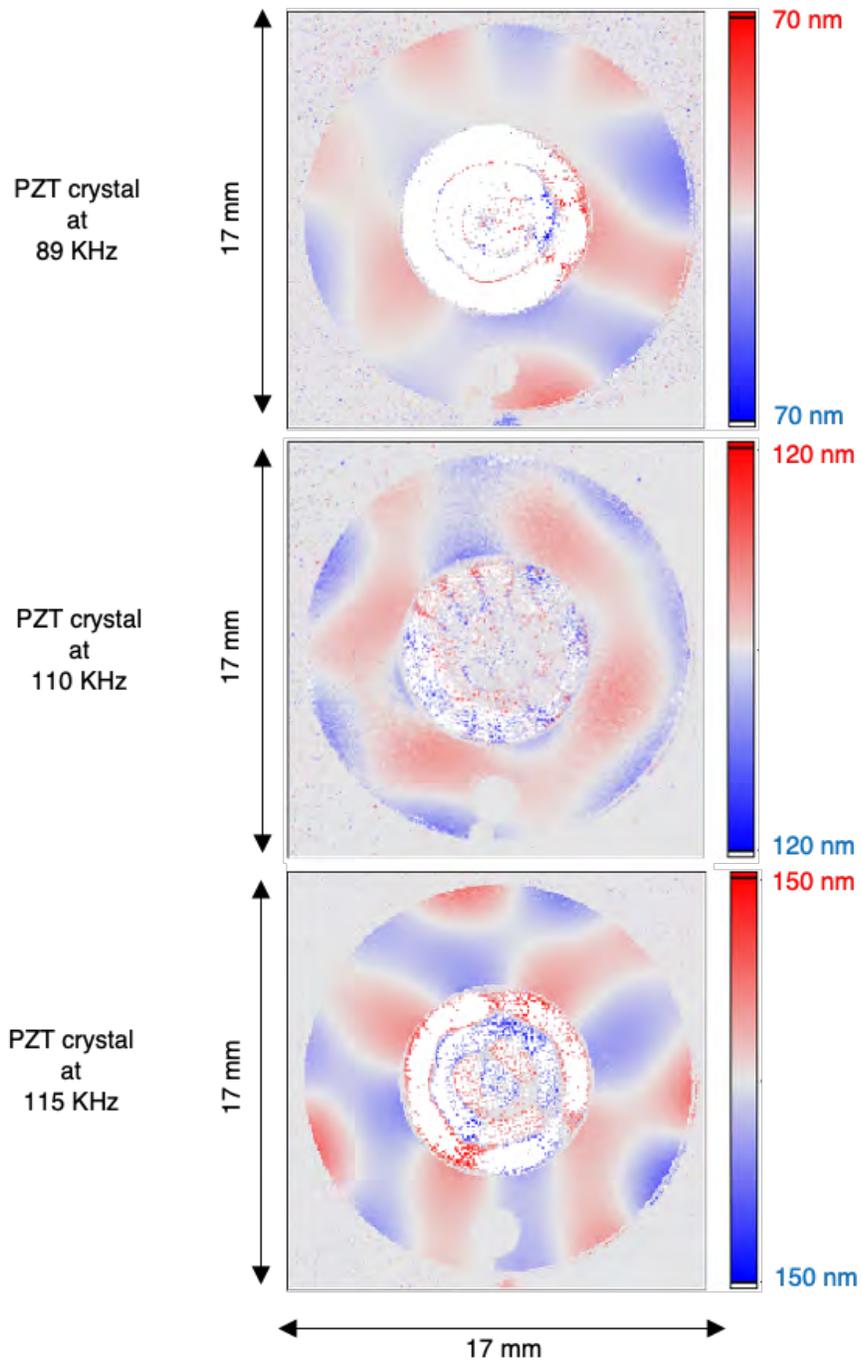

**Supplementary Figure S12. Out-of-plane global modal analysis of the entire ring PZT crystal (16 mm in diameter).** The modal analysis was conducted at three different frequencies: 89, 110, and 115 kHz (from top to bottom). The data reveal that the out-of-plane longitudinal (thickness mode) deflection of the PZT is extremely low compared to longitudinal deflection of the steel membrane. The corresponding representative animations of the out-of-plane deflection patterns of the vibrating PZT are included as Supplementary Movie 5.

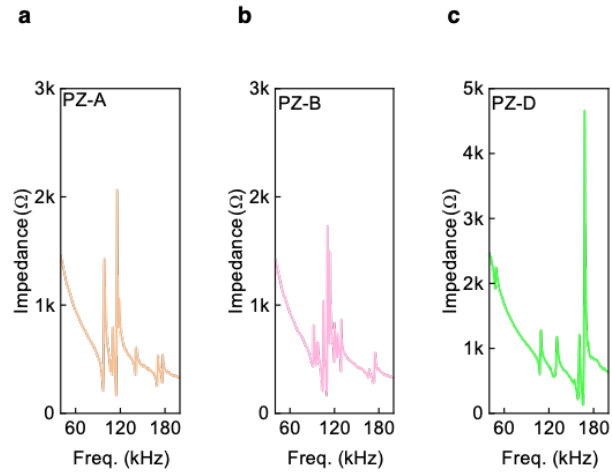

Supplementary Figure S13. Phase-impedance spectra of PZ-A, B, and D (a-c) actuators.

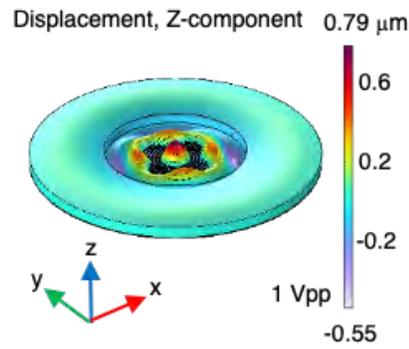

Supplementary Figure S14. Simulated deflection amplitude of a PZ-C actuator under 1 V$_{pp}$ electric stimulation.

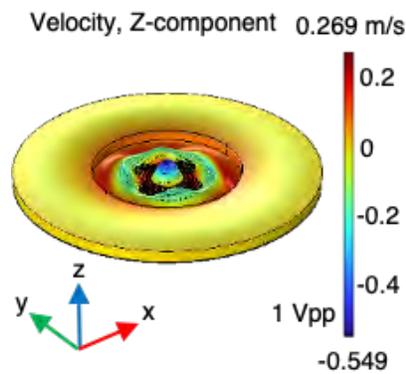

Supplementary Figure S15. Simulated velocity of a PZ-C actuator under 1 V$_{pp}$ electric stimulation.

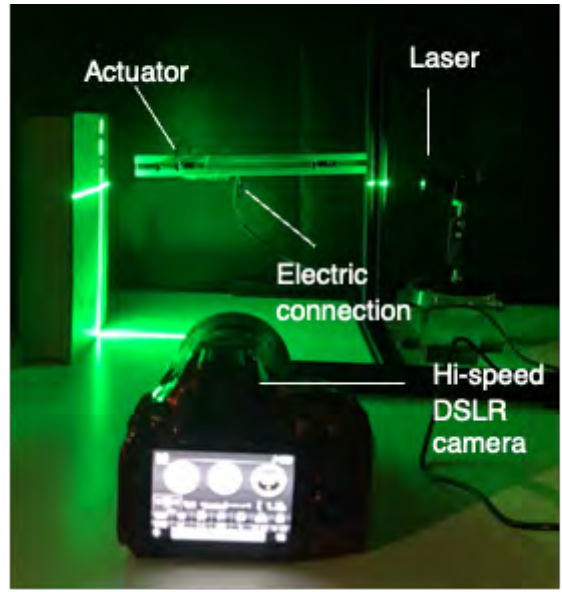

**Supplementary Figure S16. High-speed photography setup**. The setup allows capturing and recording water droplet ejection through the nozzles of piezoelectric actuator, which cannot be seen by the naked eye.

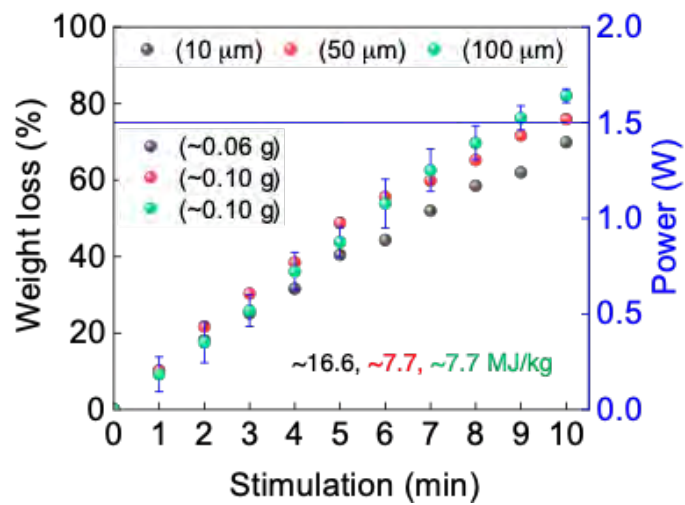

**Supplementary Figure S17. Water extraction using commercial hydrogels (HG-M) with actuators (PZs- A, B, and C) under a constant 1.5 W power.** The plot shows the desorption capabilities as a function of the nozzle size of the meshed membrane (data shown for 10, 50, and 100 μm nozzles).

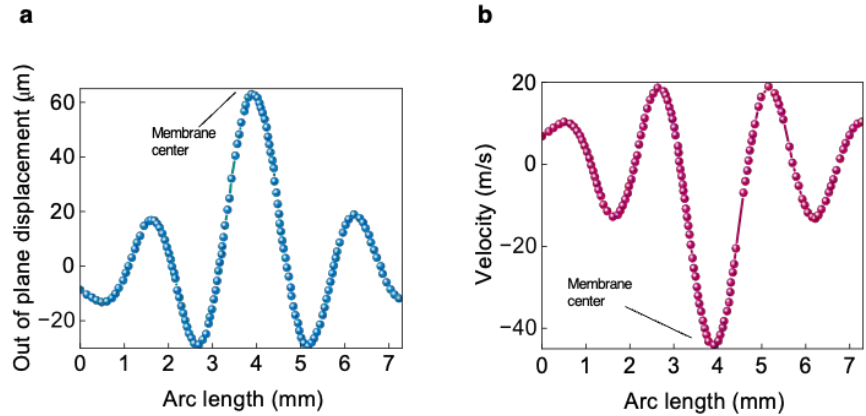

**Supplementary Figure S18. Simulating the real-life dynamic behavior of the piezoelectric actuator operating at ~114 V_pp at operating frequency of ~110 KHz. a-b.** The predicted mechanical deflection (a) and the corresponding velocity of the actuator (b) obtained from a 3D FEM model.

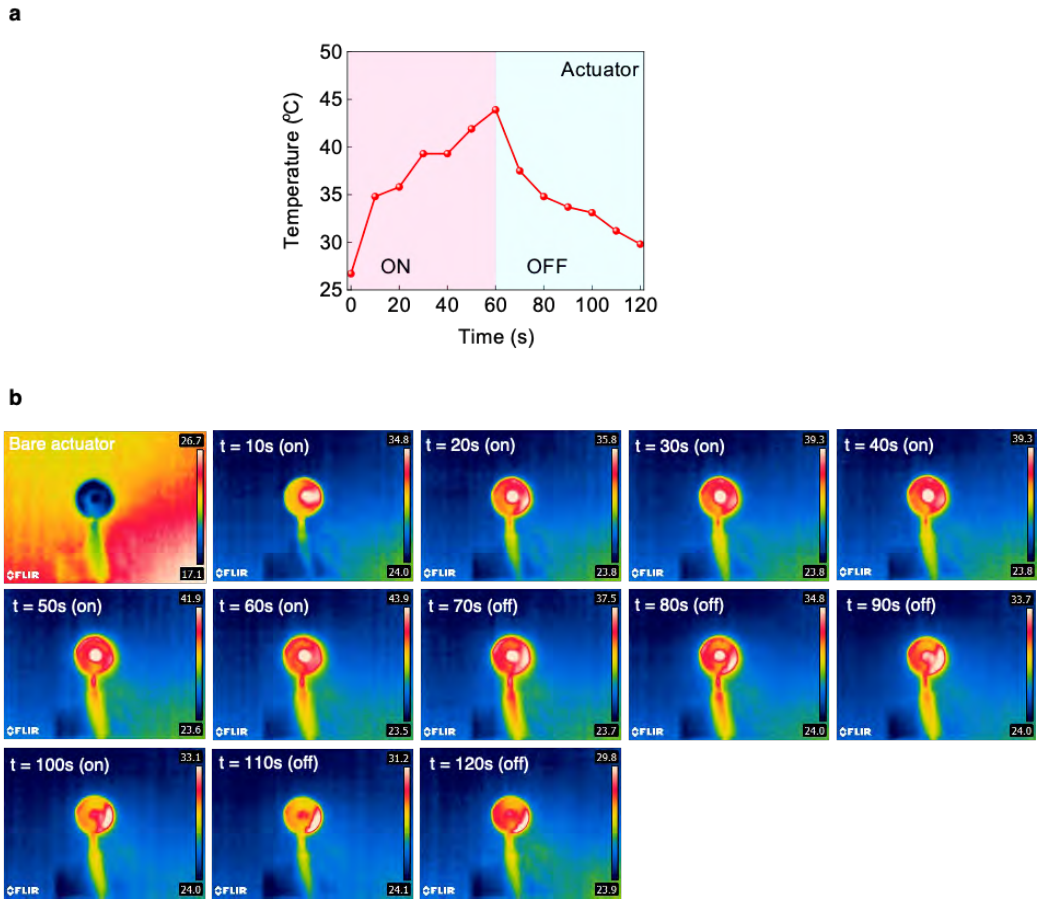

**Supplementary Figure S19. Joule heating performance of the actuator**. **a**. Time-temperature profile curve of the actuator demonstrating fast heating and cooling capability. **b**. IR thermographs acquired during the heating and cooling process. Colormap is in units of Celsius (°C).

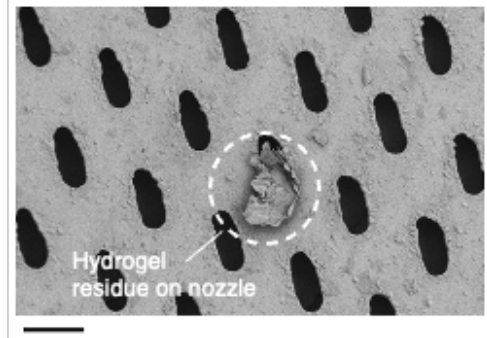

**Supplementary Figure S20. Scanning electron micrograph (SEM) of the membrane nozzle after a water extraction cycle from a hydrogel specimen.** The image shows negligible amount of hydrogel residue on the nozzle. Scale bar: 16 mm.

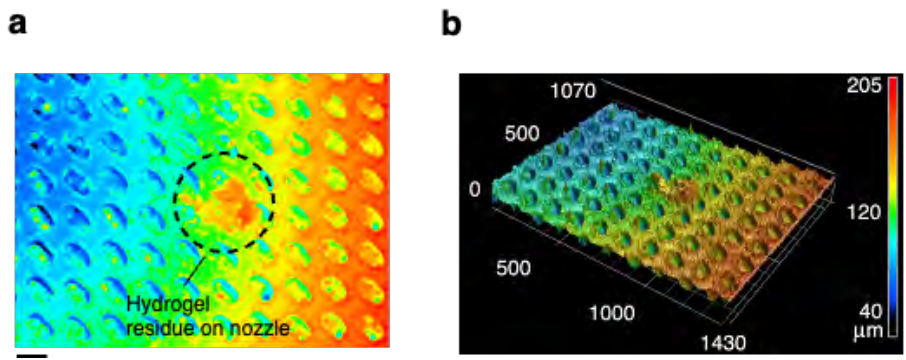

**Supplementary Figure S21. Laser scanning confocal microscopy for non-contact surface profiling of a membrane nozzle following water extraction from a hydrogel specimen. a-b.** The spatial maps of surface topography (a) and surface roughness (b). Scale bar: 100 μm (a-b).

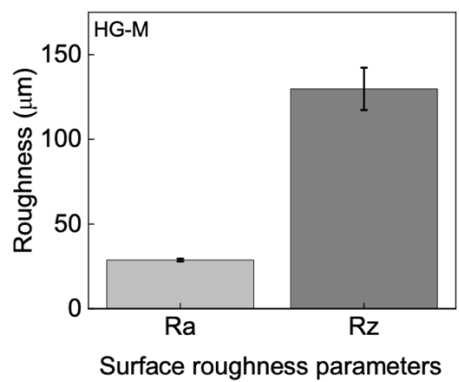

**Supplementary Figure S22. Measured roughness of the membrane nozzle after the water extraction cycle from a hydrogel specimen.** Following the assessment of multi-line roughness measurements, the arithmetic mean roughness (Ra) (i.e., average variation in profile height from the centerline) and the maximum roughness (Rz) (i.e., vertical distance between the highest and lowest points of the profile) were determined to be 28.7±0.8 and 129.8±12.5 μm, respectively.

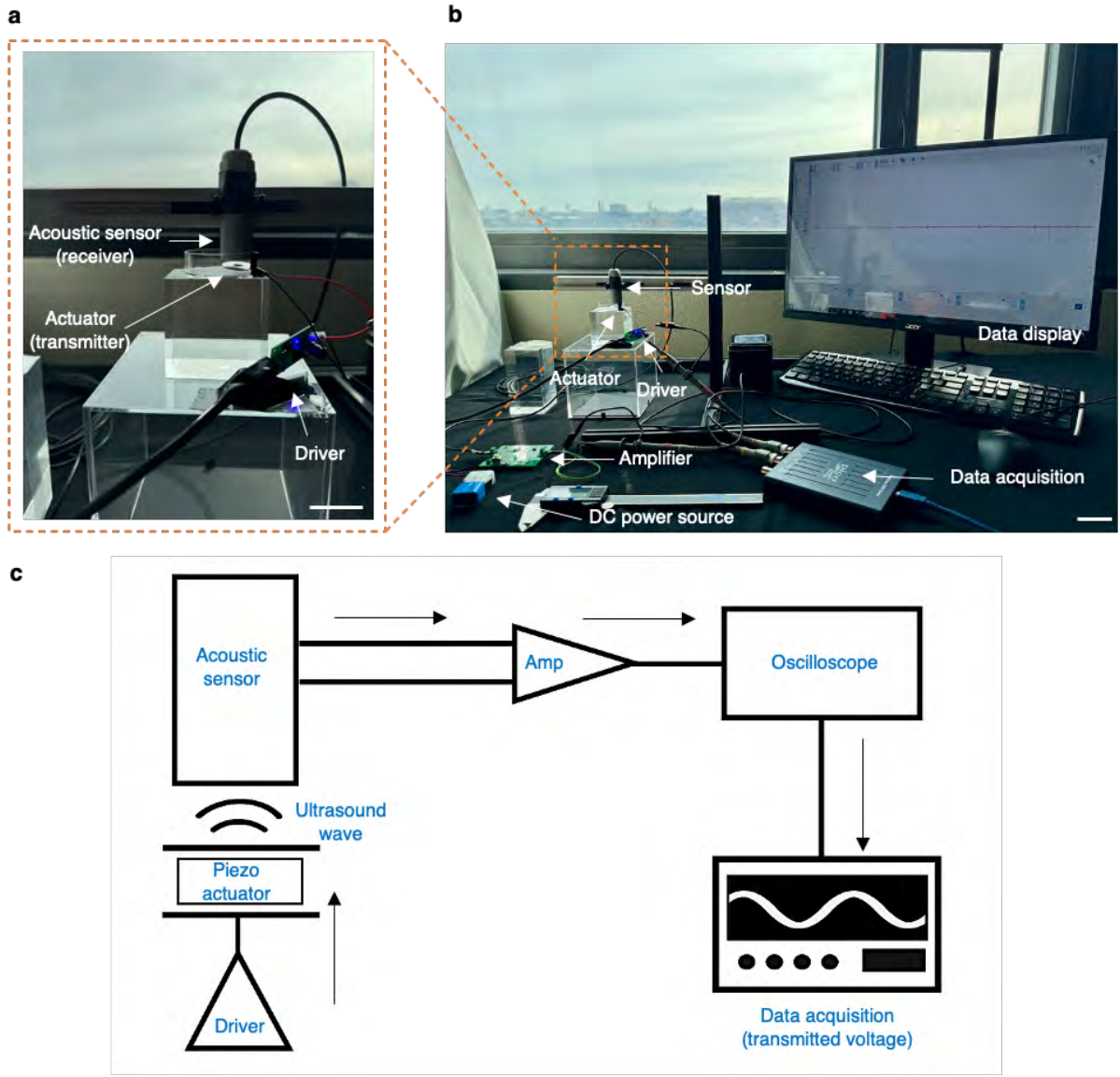

**Supplementary Figure S23. Acoustic radiation measurement test. a**. Photograph of an acoustic sensor used to measure the acoustic radiation from the piezoelectric transmitter. **b**. A photograph of the test setup showing hydrogel specimens sandwiched between the sensor and the transmitter to evaluate the impact of the hydrogel thickness and rigidity on the acoustic signal transmittance. The transmitted and received waveforms are processed for data analysis. **c**. A representative schematic diagram of the test process and the data acquisition setup. Scale bars: 16 mm (a); 52 mm (b). A representative real-time experimental demonstration of the sensor operation is included as Supplementary Movie 6.

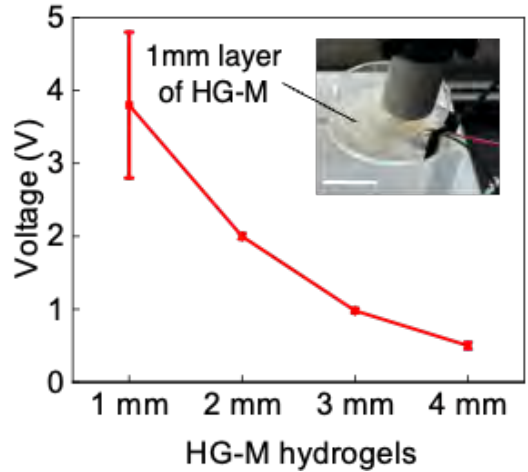

**Supplementary Figure S24. The effect of hydrogel thickness on the transmitted voltage from the piezoelectric ultrasonic actuator**. The signal is received by the acoustic hydrophone sensor (shown in the inset). Scale bar: 16 mm.

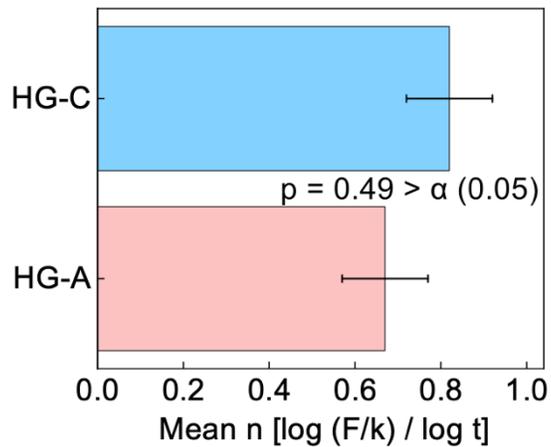

**Supplementary Figure S25. An analysis of variance (ANOVA) study to compare the release constants (n) of HG-A and HG-C hydrogel specimens** estimated using the Korsmeyer-Peppas (K-P) kinetic model.

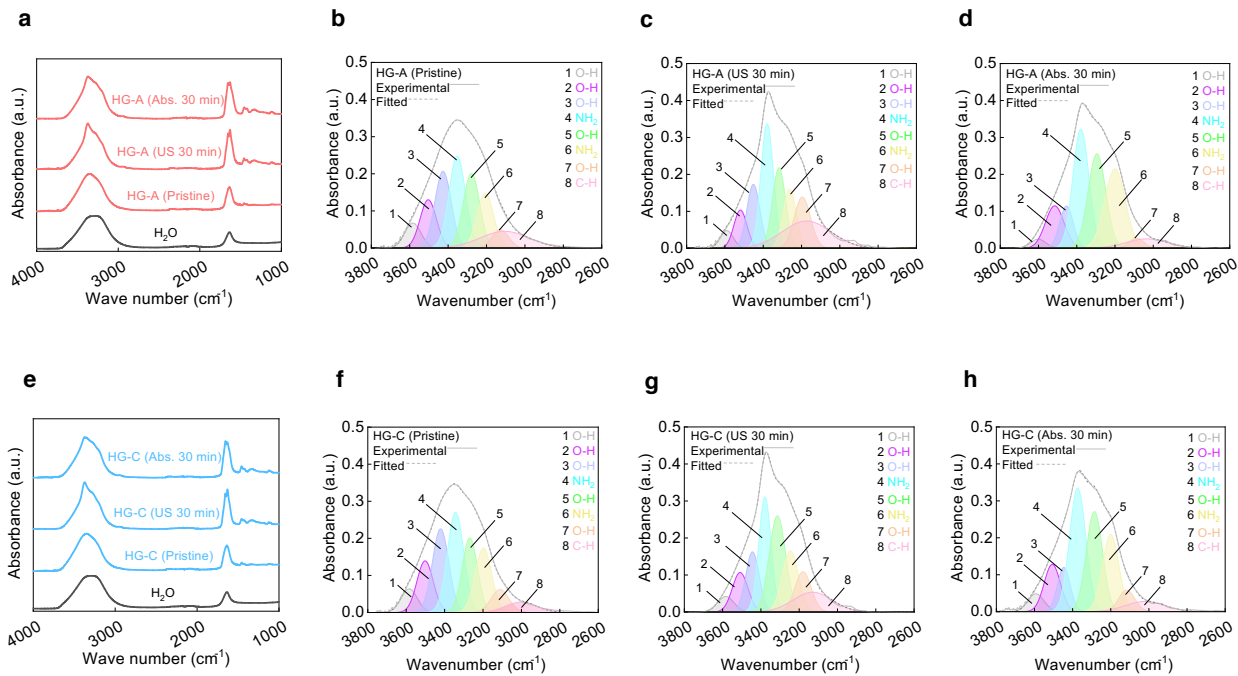

**Supplementary Figure S26. a, e.** FTIR spectra of liquid DI water, pristine HG-A and HG-C hydrogels, HG-A and HG-C subjected to 30 min of ultrasound exposure, and post-atmospheric water harvesting for 30 min following ultrasound exposure. **b-d, f-h.** Deconvoluted FTIR spectra of HG-A (b-d) and HG-C (f-g) gels during different stages of the sorption-extraction process.

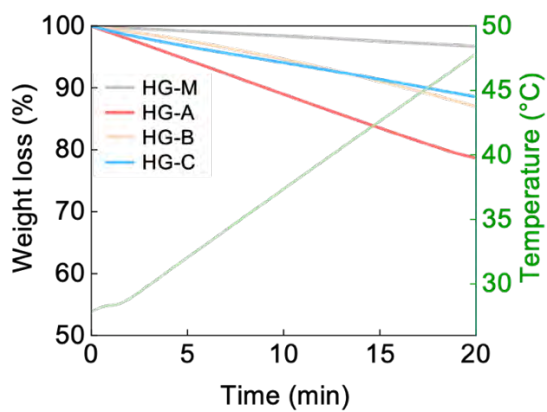

**Supplementary Figure S27. Thermogravimetric analysis of the hydrogel specimens with a temperature ramp rate of 1 °C/min.** The analysis demonstrates that the high-modulus specimens (HGs-B, C, and M) release less moisture than the low-modulus HG-A specimens.

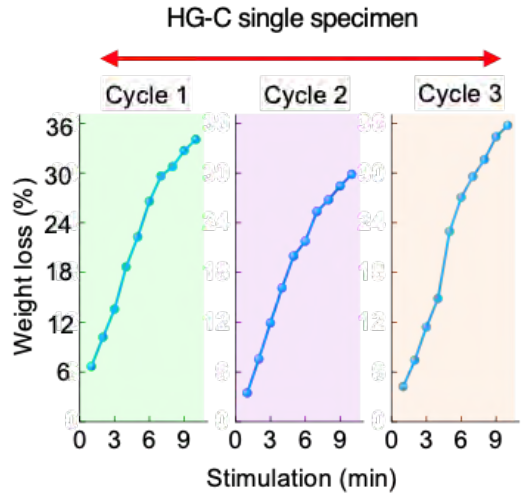

**Supplementary Figure S28. Cyclic stability test of a HG-C specimen.** The test demonstrates the desorption behavior of the same specimen over three extraction cycles using a common actuator.

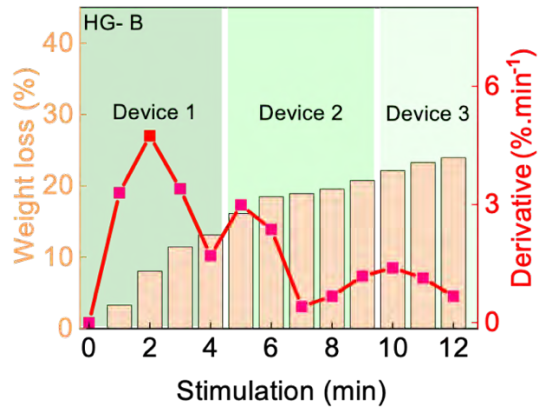

**Supplementary Figure S29. Water extraction performance from the HG-B specimen**. An erratic water extraction behavior was observed during the actuation process.

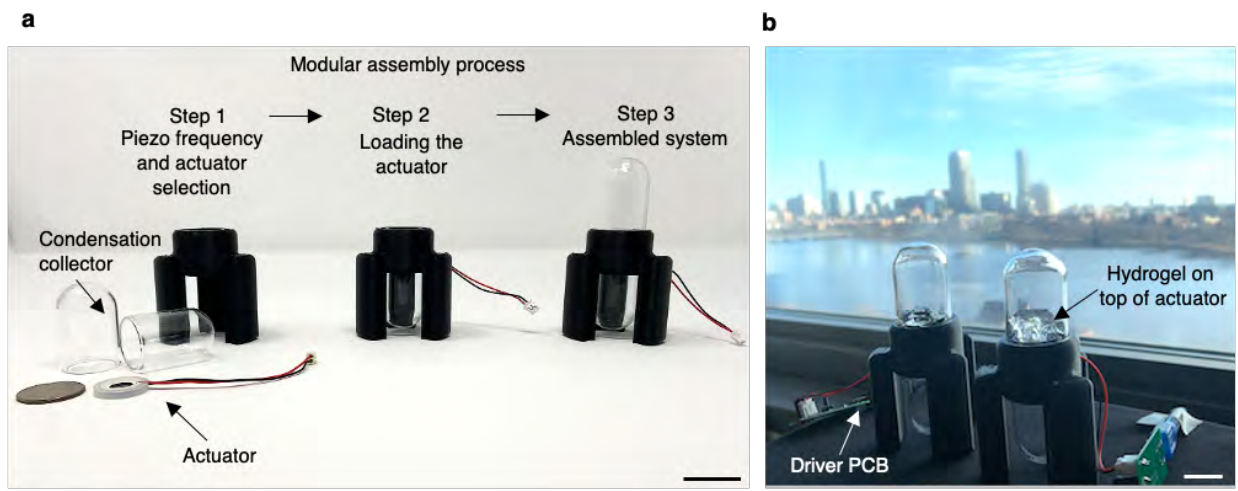

**Supplementary Figure S30. A photograph of the ultrasonic actuator system and its assembly process**. **a**. The assembly process of the entire system housing a piezoelectric actuator. **b**. A real-life indoor demonstration of the water extraction system using atmospheric water harvesting hydrogels. Scale bars: 16 cm (a-b). A representative animation and real-time demonstration of the water extraction system are included as Supplementary Movie 7.

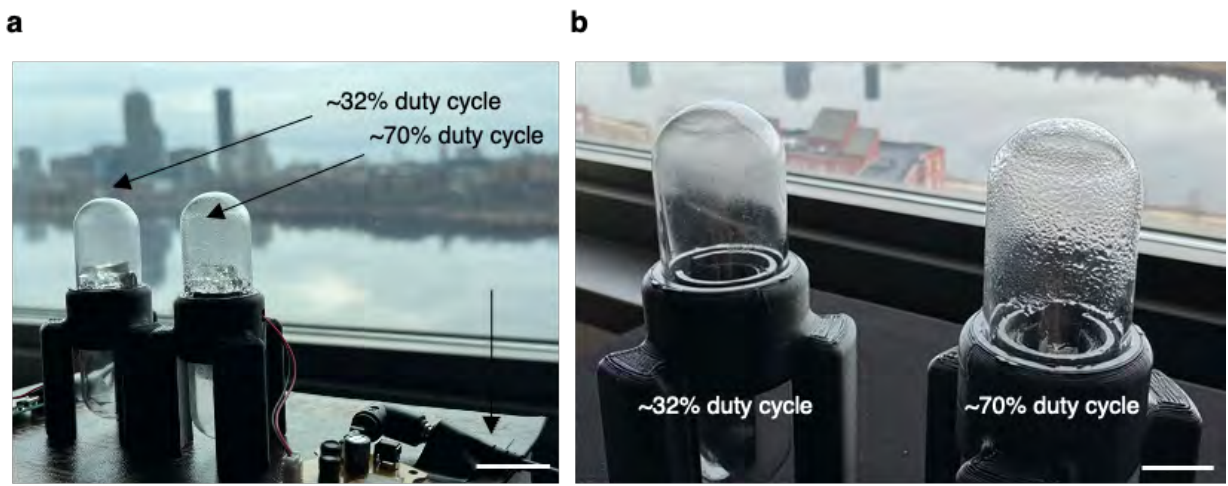

**Supplementary Figure S31. A photograph of the ultrasonic actuator system loaded with HG-C specimens**. **a**. Moisture release can be controlled by modulating the duty cycle from low (~32%) to high (~70%) as a function of time. **b**. A close view of the condensed water particles collected inside the two ultrasonic actuator systems. Scale bars: 16 cm (a); 8 cm (b).

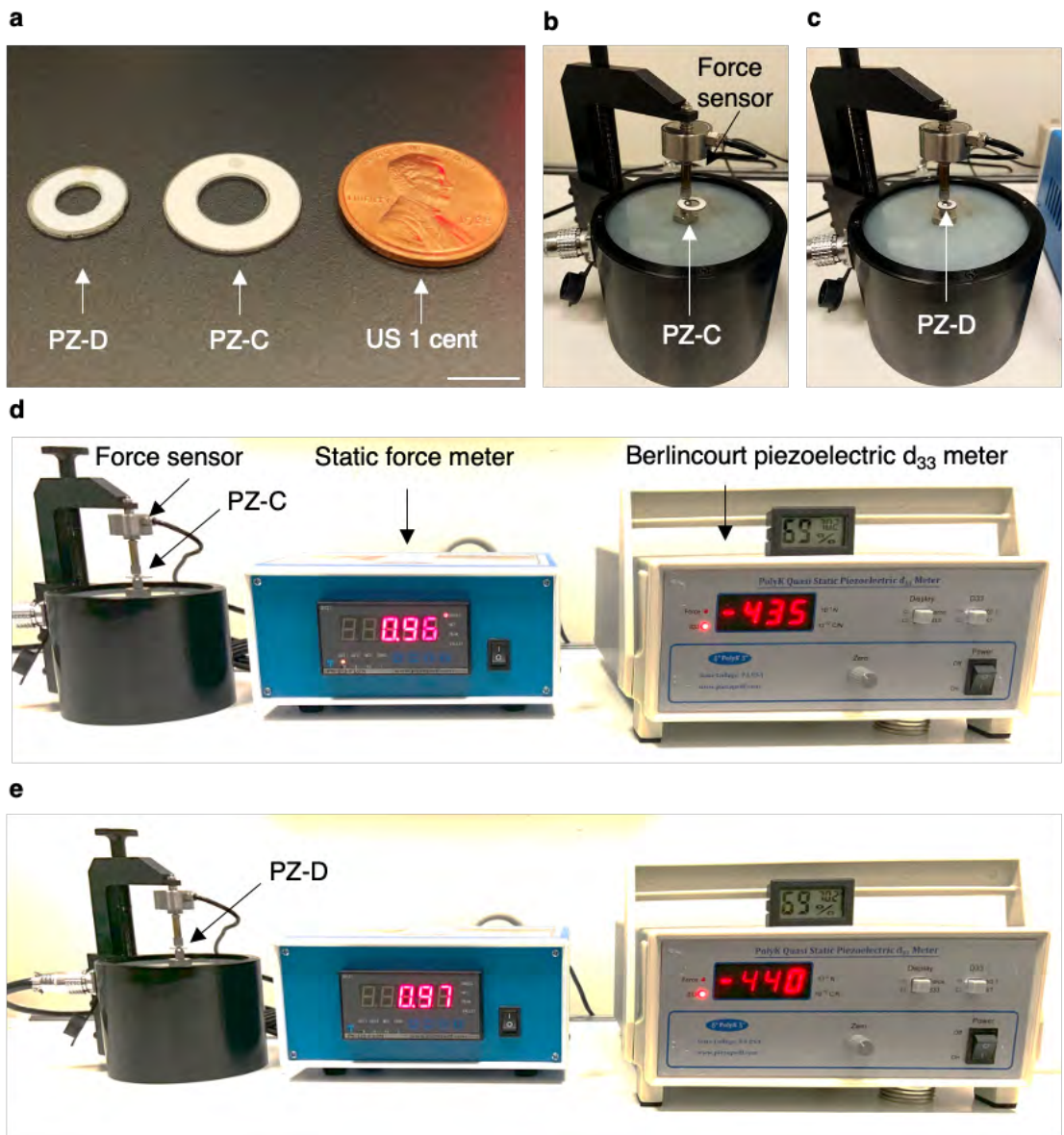

Supplementary Figure S32: Piezoelectric coefficient test setup in an ambient environment using a Berlincourt $d_{33}$ (pC/N) meter. **a** Photo of PZ-C and PZ-D piezoelectric polycrystalline ceramics used in the ultrasonic transducer devices. **b, d** Characterizing the piezoelectric constant of PZ-C (|$d_{33}$| = 435 pC/N) and **c, e** PZ-D (|$d_{33}$| = 440 pC/N) actuator devices. The tests were conducted at a constant static force of ~1N as depicted in the photos at room temperature. Scale bar, 8 mm **(a).**

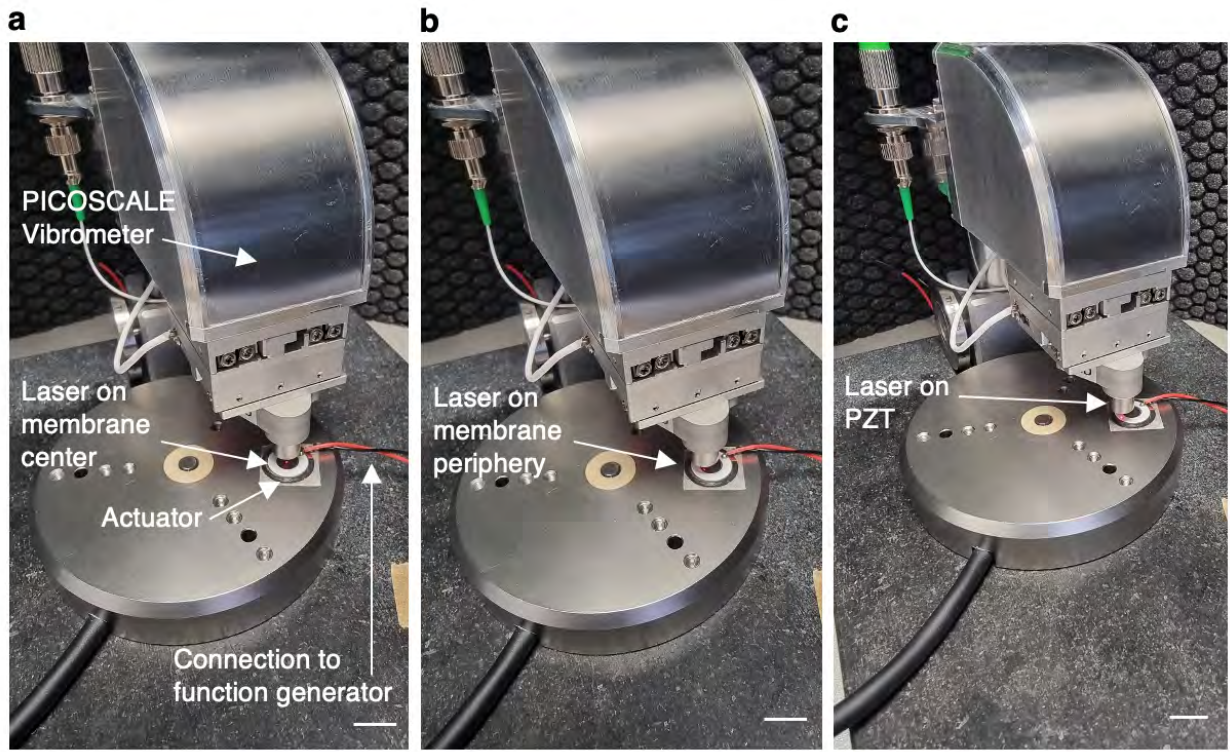

**Supplementary Figure S33: Vibration deflection analysis using a PICOSCALE vibrometer.** Laser beam focused on **a** the center of the mesh (M) and **b** a periphery (P) of the membrane. **c** Laser beam focused on the piezoelectric crystal. Scale bar, 16 mm **(a-c).**

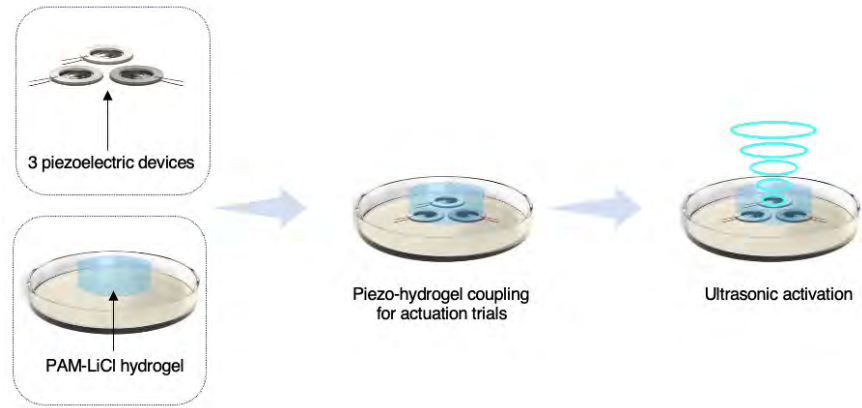

Supplementary Figure S34: Desorption test of HG-A hydrogel specimens employing three individual piezoelectric ultrasound devices together.

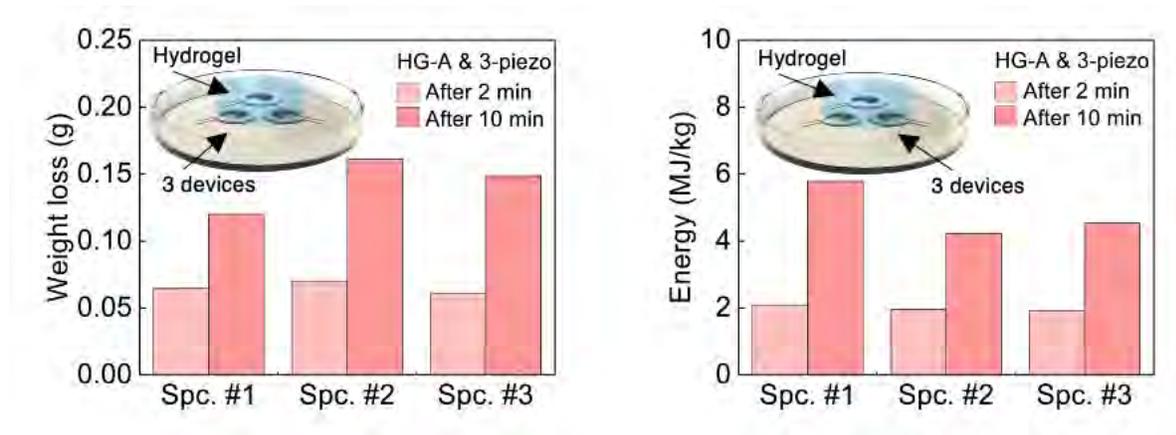

Supplementary Figure S35: Desorption behaviour of HG-A hydrogel (~2.6 g per specimen) under the actuation of three individual piezoelectric ultrasound devices for 10 minutes at 25% duty cycle. The weight loss (left) and corresponding energy consumption (right) of the specimens (~2.6 g per specimen) after 2 and 10 min, respectively.

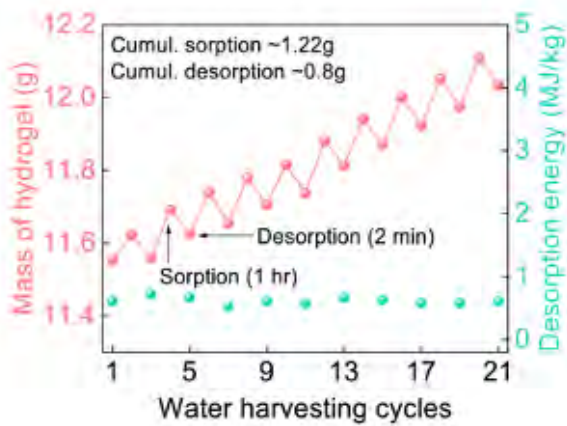
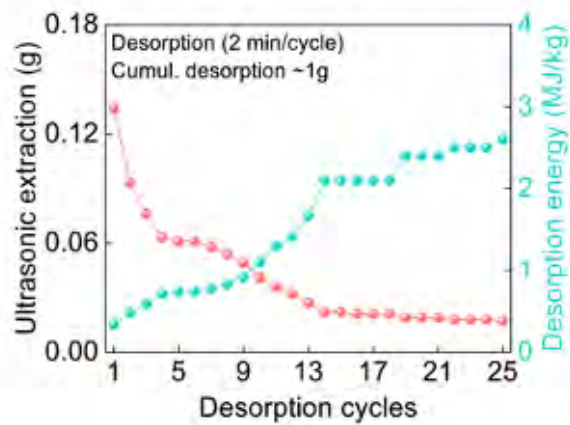

**Supplementary Figure S36: Two distinct strategies for water extraction from a sorbent: ten 1-hr sorption cycles separated by 2-min- extraction periods (a) and 25 cycles of subsequent 2-min extraction periods (b).** This shows the stability, repeatability, and energy efficiency of the proposed system using a large specimen (>16 mm diameter) for atmospheric water sorption and desorption with the energy consumption below the thermal limit.

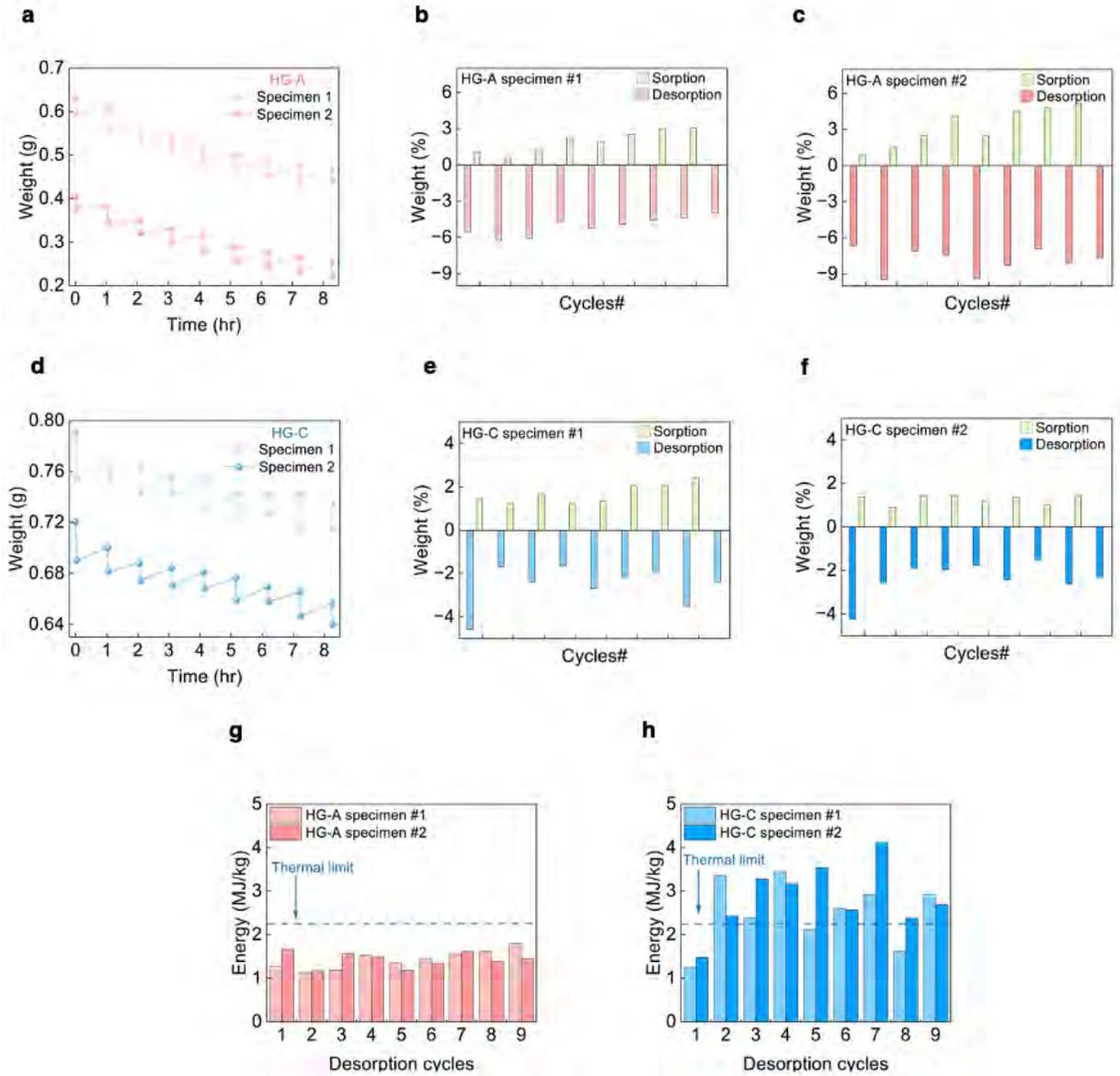

**Supplementary Figure S37: Stability, repeatability, and energy efficiency of the proposed system** using smaller HG-A specimens (≤16 mm diameter) for atmospheric water sorption and desorption with the energy consumption below the thermal limit for both HG-A (a-c, g) and HG-C (d-f, h).

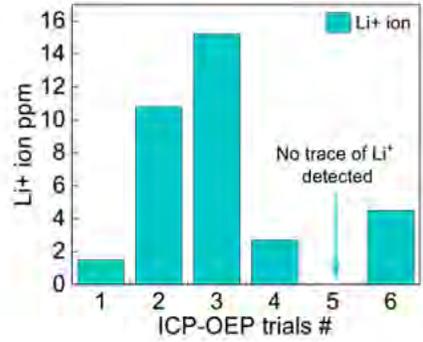

**Supplementary Figure S38:** Inductively coupled plasma optical emission spectrometry (ICP-OES) elemental analysis of the lithium-ion content in extracted water after sorption at 20% RH.

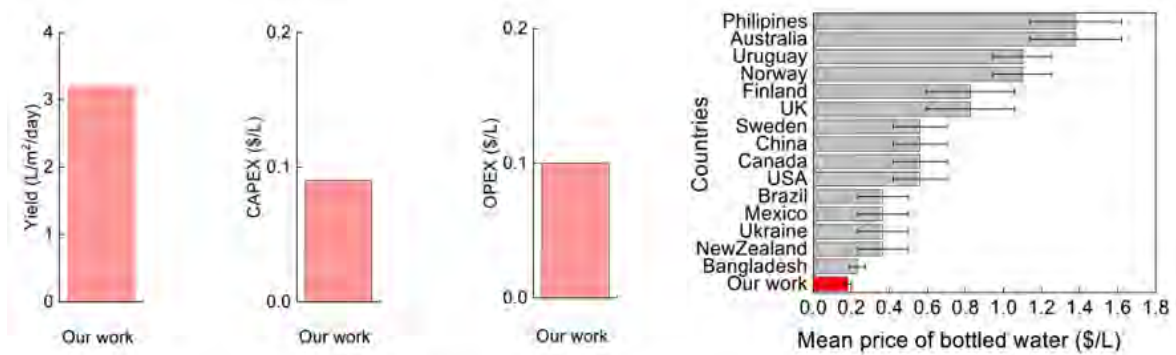

**Supplementary Figure S39:** A summary of the parameters (productivity, OPEX, and CAPEX) used in our techno-economic analysis to estimate the price of water ($/L) and a comparative study of mean price of bottled water among the countries.

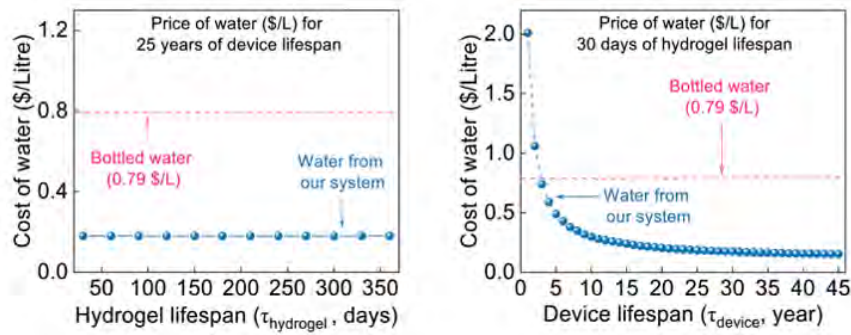

**Supplementary Figure S40**: Estimated cost of water as a function of hydrogel (a) and device (b) lifetime.

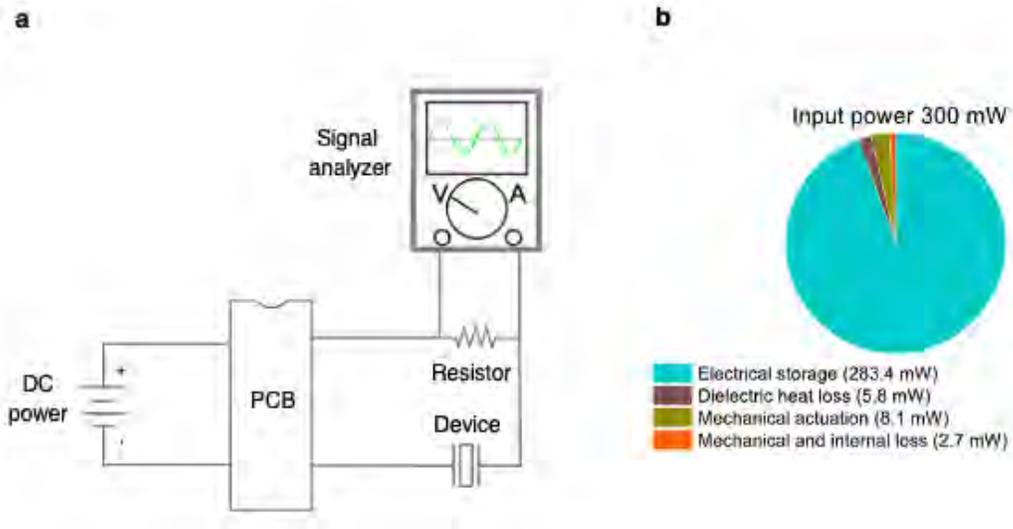

**Supplementary Figure S41:** Circuit diagram (**a**) used to characterize the input power fed to the piezoelectric transducer and a rough estimate of the power breakdown spent on thermal and mechanical actuation (**b**).

## 4. Supplementary Movies

Movie 1. Assembly and working principle of the water extraction system.

Movie 2. Real-time demonstration of the stretchability of high and low modulus hydrogels.

Movie 3. Real-time demonstration of the hydrogels harvesting water from atmosphere at extremely low humidity.

Movie 4. Stitched image reconstruction of the actuating out-of-plane porous membrane at 89, 110, and 115 kHz (longitudinal mode) using PICOSCALE vibrometer.

Movie 5. Stitched image reconstruction of the actuating out-of-plane (thickness mode) PZT piezoelectric polycrystal at 89, 110, and 115 kHz (thickness/longitudinal mode) using PICOSCALE vibrometer.

Movie 6. Real-time wireless ultrasound energy transfer through the atmospheric water harvesting hydrogels from the actuator.

Movie 7. Real-time demonstration of water extraction from the atmospheric water harvesting hydrogels using our system.